\documentclass[twocolumn,prx,aps,amsmath,amssymb,longbibliography]{revtex4-1}

\usepackage{graphicx}
\usepackage{dcolumn}
\usepackage{bm}

\begin{document}

\title{Thermoelectric fluctuations of interfering Majorana bound states}

\author{Sergey Smirnov}
\affiliation{P. N. Lebedev Physical Institute of the Russian Academy of
  Sciences, 119991 Moscow, Russia}
\email{1) sergej.physik@gmail.com\\2)
  sergey.smirnov@physik.uni-regensburg.de\\3) ssmirnov@sci.lebedev.ru}

\date{\today}

\begin{abstract}
Nonequilibrium states produced by electric and thermal voltages ($V$, $V_T$)
provide a straightforward insight into underlying degrees of freedom of
composite nanostructures and are of particular interest to probe Majorana
bound states. Here we explore fluctuations of thermoelectric currents at
finite frequencies $\omega$ in a quantum dot coupled to two interfering
Majorana bound states. At small $V$ we find that in the emission spectra the
differential thermoelectric quantum noise $\partial S^>/\partial V_T$ shows an
antiresonance whereas in the absorption spectra Majorana interference induces
an antiresonance-resonance pair. At large $V$ this pair is preserved whereas
the emission antiresonance turns into an antiresonance-resonance pair
identical to the absorption one making $\partial S^>/\partial V_T$
antisymmetric in the frequency $\omega$. This antisymmetry distinguishes
Majorana behavior from the one induced by Andreev bound states and does not
break at higher temperatures making it attractive for experiments on Majorana
interference via thermoelectric fluctuation response.
\end{abstract}

\maketitle

\section{Introduction}\label{intro}
Probing Majorana bound states (MBSs) in diverse systems \cite{Kitaev_2001,
Alicea_2012,Flensberg_2012,Sato_2016,Aguado_2017,Lutchyn_2018,Tsintzis_2022,
Marra_2022,Dvir_2023,Muralidharan_2023,Tsintzis_2024,Tanaka_2024} via various
advanced techniques enriches our knowledge about many remarkable aspects of
Majorana nanosystems and creates a significant potential for future
experimental implementations. Involving a complex interplay between various
degrees of freedom such nanosystems, on one side, are appealing for quantum
computations \cite{Kitaev_2003}, based on the unique non-Abelian exchange
statistics of MBSs (as opposed to the Abelian one of the Majorana fermions
\cite{Majorana_1937,Itzykson_zuber_1980} in the particle physics), and, on the
other side, they exhibit exceptional physical properties which may be accessed
by proper measurements. The latter side of exploring Majorana nanosystems
offers a great freedom to explore MBSs via numerous and essentially
independent physical observables which may be predicted in theory and would be
relevant for state-of-the-art experiments.

For example, equilibrium states of nanoscopic devices involving MBSs may be
addressed via their linear conductance or entropy which provide quantum
transport or thermodynamic probes, respectively. These probes are
fundamentally distinct. The most essential difference between them, among
others, manifests in whether the corresponding outcomes admit unambiguous
conclusions. Whereas the fractional Majorana entropy
\cite{Smirnov_2015,Sela_2019,Smirnov_2021,Smirnov_2021a} cannot be ascribed to
non-Majorana quasiparticles, the values of various linear conductances,
predicted for systems with MBSs, may also be obtained in experiments where
MBSs are absent \cite{Yu_2021,Frolov_2021}. Nevertheless, the advantage of the
quantum transport approach consists in its technological flexibility which
allows one to straightforwardly apply it to various systems with MBSs
\cite{Mourik_2012,Nadj-Perge_2014,Wang_2022}. Quantum thermodynamic
measurements of the entropy in nanoscopic systems with MBSs are currently an
experimental challenge whose solution appears quite plausible as one may
assume on the basis of the notable progress \cite{Hartman_2018,Kleeorin_2019,
Pyurbeeva_2021,Child_2022,Han_2022,Pyurbeeva_2022,Child_2022a} in theoretical
and experimental research devoted to the entropy of nanoscopic systems whose
degrees of freedom do not yet involve MBSs.

Due to its experimental appeal quantum transport is of special interest and is
often used to analyze MBSs not only via linear conductances but also via
differential conductances, or in general, via mean electric currents beyond
the linear response approach \cite{Liu_2011,Fidkowski_2012,Prada_2012,
Pientka_2012,Lin_2012,Lee_2013,Kundu_2013,Vernek_2014,Ilan_2014,Lobos_2015,
Peng_2015,Sharma_2016,Heck_2016,Das_Sarma_2016,Lutchyn_2017,Weymann_2017,
Campo_Jr_2017,Liu_2017,Huang_2017,Liu_2018,Lai_2019,Tang_2020,Zhang_2020,
Chi_2021,Galambos_2022,Jin_2022,Zou_2023,Huguet_2023,Becerra_2023,Ziesen_2023,
Yao_2023,Taranko_2024}. When Majorana nanosystems are coupled to
contacts with different temperatures, one gets an opportunity to study
thermoelectric mean currents \cite{Leijnse_2014,Lopez_2014,Khim_2015,
Ramos-Andrade_2016,Ricco_2018,Smirnov_2020a,Wang_2021,He_2021,Giuliano_2022,
Buccheri_2022,Bondyopadhaya_2022,Zou_2022,Wang_2023,Zou_2023a,Chi_2024,
Mishra_2024,Trocha_2025}
providing information independent of pure electric behavior of MBSs. In fact,
strongly nonequilibrium Majorana systems beyond the linear response regime
provide a wider spectrum of measurements which are not restricted only by mean
currents. When the fluctuation-dissipation theorem is not applicable, as is
the case in strong nonequilibrium, electric and thermoelectric fluctuations
bring information unavailable in conductance measurements and proper
current-current correlation functions allow one to analyze MBSs from a
qualitatively different side. Indeed, one may additionally investigate MBSs
via the electric \cite{Liu_2015,Liu_2015a,Haim_2015,Valentini_2016,
Zazunov_2016,Smirnov_2017,Jonckheere_2019,Bathellier_2019,Smirnov_2019,
Manousakis_2020,Smirnov_2022,Feng_2022,Cao_2023,Smirnov_2024}
and thermoelectric \cite{Smirnov_2018,Smirnov_2019a,Smirnov_2023} shot and
quantum noise which are physical observables of experimental relevance. Shot
noise experiments have already started dealing with Majorana nanosystems
\cite{Ge_2023}. Experimental measurements of the differential quantum noise at
finite frequencies have been previously performed \cite{Basset_2012} to study
the Kondo effect in a quantum dot (QD) and one may naturally assume that the
techniques used for these measurements could be adapted to QDs coupled to
MBSs. It is important to note that this coupling often involves Majorana
tunneling phases which, on one side, are important degrees of freedom to drive
Majorana qubits \cite{Gau_2020,Gau_2020a} and, on the other side, are
responsible for various interference effects resulting in a remarkable
behavior of the shot and quantum noise  \cite{Smirnov_2022,Smirnov_2024}.

Various important physical observables characterizing random deviations of
currents from their mean values, such as the static shot noise or finite
frequency quantum noise, exhibit a rather nontrivial behavior in
nonequilibrium states produced by both bias and thermal voltages
\cite{Smirnov_2018,Smirnov_2019a,Smirnov_2023}. In particular, in the absence
of Majorana interference effects, the static (zero-frequency) differential
thermoelectric shot noise has been investigated in Ref. \cite{Smirnov_2018},
the differential thermoelectric quantum noise at finite frequencies has been
presented in Ref. \cite{Smirnov_2019a} and the zero-frequency differential
shot noise, more specifically its crossover behavior, as a function of the
thermal voltage has been explored in Ref. \cite{Smirnov_2023}. In the presence
of Majorana interference effects but in the absence of thermal voltages, that
is pure electric behavior of the zero-frequency differential shot noise and
finite frequency differential quantum noise, has been analyzed in,
respectively, Refs. \cite{Smirnov_2022} and \cite{Smirnov_2024}. More
interesting phenomena are expected when Majorana interference effects emerge
in nanoscopic systems driven by both electric and thermal voltages. When the
competition between electrically and thermally excited quasiparticle flows is
additionally boosted by Majorana interference effects, nonequilibrium states
resulting from this complex interplay acquire an intricate nature. As such,
they represent an original source of information about a unique fluctuation
response of interfering MBSs, particularly, at finite frequencies. These
fluctuations may be probed via quantum transport techniques measuring, for
example, such a physical observable as the finite frequency quantum noise
which, to our knowledge, has never been addressed for interfering MBSs in
thermoelectric nonequilibrium.

In this work we consider a QD coupled to two MBSs and explore the behavior of
the differential thermoelectric quantum noise
$\partial S^>(\omega,V,V_T,\Delta\phi)/\partial V_T$ defined as the derivative
of the greater noise correlation function with respect to the thermal voltage
$V_T$. The analysis is performed numerically at finite frequencies $\omega$
for both the emission ($\omega<0$) and absorption ($\omega>0$) parts of the
spectra and focuses particularly on the behavior of the interfering MBSs, that
is on the one arising for finite values of the Majorana tunneling phase
difference $\Delta\phi$. In the regime of small bias voltages $V$ it is shown
that for well separated energy scales there develops an antiresonance in the
emission part of the spectra and that it is located around
$\hbar\omega=-|eV|/2$. This emission antiresonance results from the Majorana
interference because it is strongly suppressed for $\Delta\phi=0$. In the
absorption part of the spectra there develops a pair antiresonance-resonance
located around $\hbar\omega=|eV|/2$. Like the emission antiresonance, this
pair is of pure Majorana interference nature since it is also strongly
suppressed for $\Delta\phi=0$. Thus in the regime of small bias voltages and
for $\Delta\phi\neq 0$ the emission and absorption spectra are not related to
each other. This is in contrast to the case where a QD is effectively coupled
to a single MBS and the emission and absorption spectra turn out to be
symmetrically related in the regime of small bias voltages
\cite{Smirnov_2019a}. Further, it is found that the absorption
antiresonance-resonance pair located at $\hbar\omega=|eV|/2$ survives in the
regime of large bias voltages but with the exactly twice reduced amplitudes of
the antiresonance and resonance. At the same time the emission antiresonance
is fully destroyed by large bias voltages and transforms into an emission
antiresonance-resonance pair located at $\hbar\omega=-|eV|/2$. This emission
antiresonance-resonance pair is fully identical to the absorption one for any
finite value of $\Delta\phi$. This transformation of the emission spectra at
large bias voltages is of particular importance. Indeed, since the
differential thermoelectric quantum noise is strongly suppressed outside
vicinities of the frequencies $\hbar\omega=\pm|eV|/2$, we find that
$\partial S^>(-\omega,V,V_T,\Delta\phi)/\partial V_T=
-\partial S^>(\omega,V,V_T,\Delta\phi)/\partial V_T$, that is the absorption
and emission spectra are antisymmetrically related in the regime of large bias
voltages for any finite value of $\Delta\phi$. Finally, we investigate the
regime of high bias voltages and temperatures which might be of experimental
relevance. Here, our estimate shows that the temperature may be around
$60\,\text{mK}$ (see the details at the end of Section \ref{concl} with the
notations and parameter regime defined in the main text) which is well within
the temperature range used for quantum transport experiments. It is shown that
the antisymmetric relation between the absorption and emission spectra is also
observed in this regime in the presence of the Majorana interference
($\Delta\phi\neq 0$). In contrast, we demonstrate that Andreev bound states
(ABSs) break this relation,
$\partial S^>(-\omega,V,V_T,\Delta\phi)/\partial V_T\neq
-\partial S^>(\omega,V,V_T,\Delta\phi)/\partial V_T$,
and their thermoelectric fluctuation response at finite frequencies $\omega$
is qualitatively different from the one of interfering MBSs.

The paper is organized as follows. In Section \ref{Ham} we specify the system
via its Hamiltonian involving tunneling between a QD and two MBSs which may
interfere. Thermoelectric nonequilibrium induced by coupling of the QD to
contacts with different chemical potentials and temperatures is also specified
here. To describe these nonequilibrium states we apply the Keldysh technique
in Section \ref{Keld}. The results obtained for the differential
thermoelectric quantum noise are presented in Section \ref{Results} in various
regimes including low bias voltages, large bias voltages and also large
temperatures in the presence of the Majorana interference. Finally, we
conclude the paper in Section \ref{concl} where we also estimate the values of
the temperature at which experiments could be performed to reveal Majorana
interference patterns in the differential thermoelectric quantum noise at
finite frequencies.
\section{Hamiltonian of the system with interfering Majorana bound states}\label{Ham}
Let us start with the full Hamiltonian of a system whose behavior is
essentially determined by Majorana interference effects:
\begin{equation}
  \hat{H}=\hat{H}_\text{QD}+\hat{H}_\text{C}+\hat{H}_\text{TS}+\hat{H}_\text{QD-C}+\hat{H}_\text{QD-TS}.
  \label{Full_Ham}
\end{equation}
The system consists of a QD, metallic contacts, and a topological
superconductor (TS) which are modeled by the Hamiltonians $\hat{H}_\text{QD}$,
$\hat{H}_\text{C}$ and $\hat{H}_\text{TS}$, respectively.

The Hamiltonian of the isolated QD,
\begin{equation}
  \hat{H}_\text{QD}=\epsilon_dd^\dagger d,
  \label{QD_Ham}
\end{equation}
represents an electronic system with one non-degenerate single-particle energy
level. Its energy $\epsilon_d$ relative to the equilibrium chemical potential
$\mu$ may be tuned via a proper gate voltage. The fermionic operators
corresponding to the single-particle state of the QD, $d$ and $d^\dagger$,
obey the conventional anticommutation relations,
\begin{equation}
  \{d,d\}=0,\quad\{d,d^\dagger\}=1.
  \label{QD_operators}
\end{equation}
As described below, the QD is coupled via tunneling to a closely located TS
(see, {\it e.g.}, Ref. \cite{Deng_2018} for a technological
implementation). The topological phase with two MBSs at the ends of the TS
emerges in a strong magnetic field which also makes the energy spectrum of the
QD non-degenerate. Indeed, computations \cite{Tijerina_2015} based on the
numerical renormalization group indicate that in a strong magnetic field the
QD behaves as the one where the spin degeneracy is removed and the Kondo
effect is fully absent. Thus the Majorana physics is properly captured
by the non-degenerate QD model in Eq. (\ref{QD_Ham}) when it is combined with
a suitable model of the TS coupled to the QD (see also
Ref. \cite{Flensberg_2011}).
\begin{figure}
  \includegraphics[width=8.0 cm]{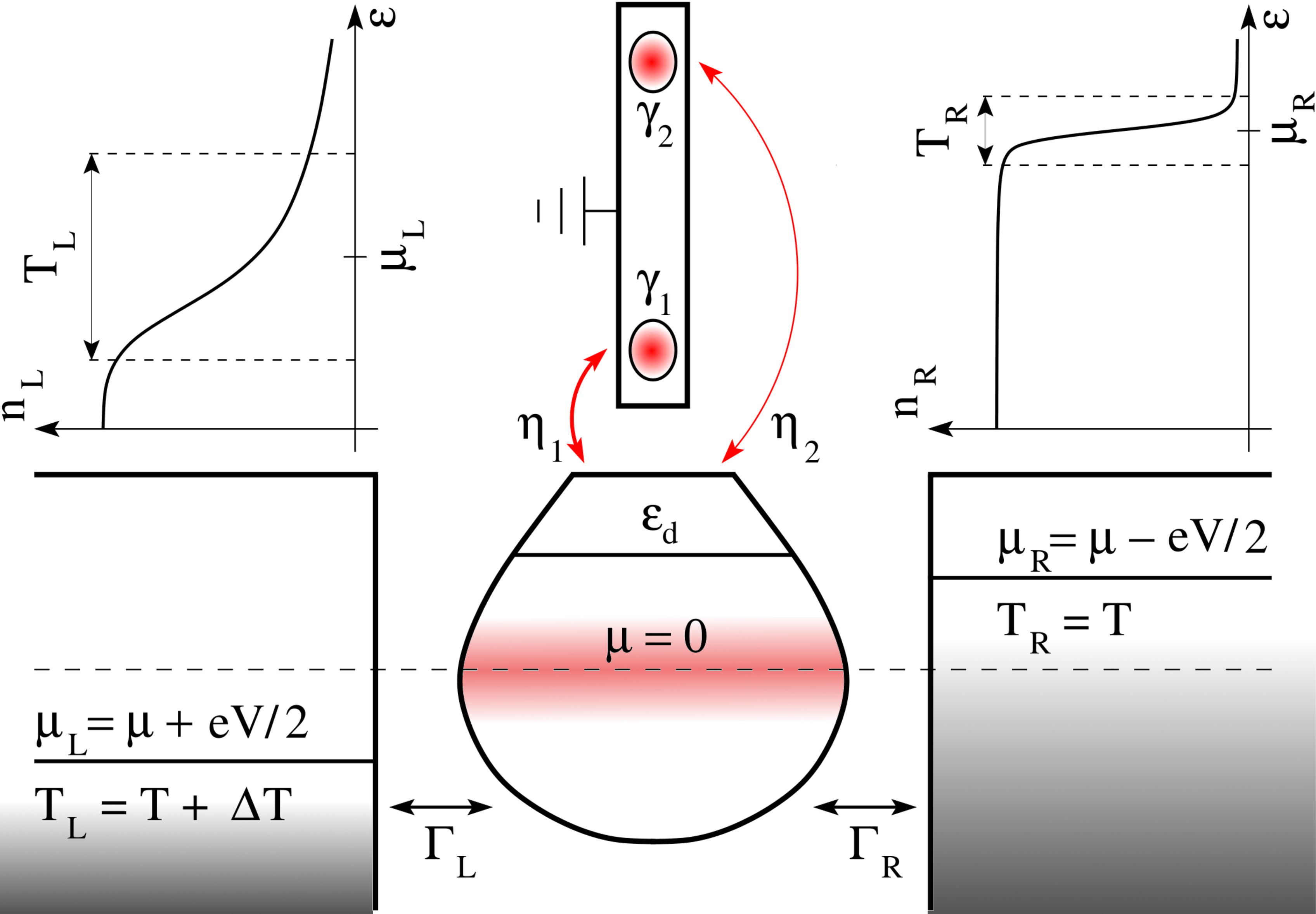}
  \caption{\label{figure_1} An illustration of the physical system modeled by
    the full Hamiltonian in Eq. (\ref{Full_Ham}). The QD with the energy level
    $\epsilon_d$ is coupled with the coupling strengths $\Gamma_{L,R}$ to,
    respectively, the left and right contacts characterized by the chemical
    potentials $\mu_{L,R}$ and temperatures $T_{L,R}$. The grounded TS
    supports two MBSs $\gamma_{1,2}$ coupled to the QD. This coupling is
    characterized by the tunneling matrix elements $\eta_{1,2}$ with the
    amplitudes $|\eta_1|\gg|\eta_2|$ modeling an asymmetric location of the
    Majorana modes with respect to the QD (see, for example,
    Ref. \cite{Deng_2018}).}
\end{figure}

The Hamiltonian of the two, left ($L$) and right ($R$), isolated metallic
contacts,
\begin{equation}
  \hat{H}_\text{C}=\sum_{l=\{L,R\}}\sum_k\epsilon_kc^\dagger_{lk}c_{lk},
  \label{Cont_Ham}
\end{equation}
describes non-interacting fermionic systems. The operators corresponding to
the states of these systems also obey the conventional anticommutation
relations,
\begin{equation}
  \{c_{lk},c_{l'k'}\}=0,\quad\{c_{lk},c^\dagger_{l'k'}\}=\delta_{ll'}\delta_{kk'}.
  \label{Cont_operators}
\end{equation}
The contacts are assumed to be massive so that their energy spectra,
$\epsilon_{L,k}=\epsilon_{R,k}=\epsilon_k$, are continuous. They are
characterized by the corresponding density of states $\nu(\epsilon)$ which is,
in fact, sufficient to calculate various physical observables. We neglect the
energy dependence of the contacts density of states,
\begin{equation}
  \nu(\epsilon)\approx\frac{1}{2}\nu_C.
  \label{DOS_cont_const}
\end{equation}
This assumption is often justified since in many quantum transport experiments
measurements are performed in the energy range where the density of states in
the metallic contacts does not vary too much to produce qualitative observable
effects.

Both metallic contacts are in equilibrium states. These states are specified
by the Fermi-Dirac distributions:
\begin{equation}
  n_{L,R}(\epsilon)=\frac{1}{\exp(\frac{\epsilon-\mu_{L,R}}{k_\text{B}T_{L,R}})+1},
  \label{F-D}
\end{equation}
where $\mu_{L,R}$ is the chemical potential and $T_{L,R}$ the temperature of
the left or right metallic contact, respectively. In general the equilibrium
states of the left and right contacts are different, that is $\mu_L\neq\mu_R$
and $T_L\neq T_R$. The difference between the chemical potentials is specified
via the bias voltage $V$ applied to the two contacts:
\begin{equation}
  \mu_{L,R}=\mu\pm eV/2,
  \label{Chem_pot_L_R}
\end{equation}
where we have used $eV<0$ to obtain the numerical results discussed in
Sec. \ref{Results}. The temperatures are chosen to specify the left contact as
hot and the right one as cold,
\begin{equation}
  T_L=T+\Delta T,\quad T_R=T,\quad\Delta T\geqslant 0.
  \label{Temp_L_R}
\end{equation}
To obtain the numerical results presented in Sec. \ref{Results} we have
parameterized the temperature difference $\Delta T$ by a thermal voltage $V_T$
between the contacts:
\begin{equation}
  eV_T\equiv k_\text{B}\Delta T.
  \label{Therm_volt}
\end{equation}

The low-energy Hamiltonian of the isolated TS,
\begin{equation}
  \hat{H}_\text{TS}=\frac{i}{2}\xi\gamma_2\gamma_1,
  \label{TS_Ham}
\end{equation}
assumes that in its topological phase there arise two MBSs located at its
ends. Here the self-adjoint fermionic operators,
\begin{equation}
  \gamma_{1,2}^\dagger=\gamma_{1,2},
  \label{TS_operators_1}
\end{equation}
corresponding to the low-energy states of the TS obey the Majorana, or
Clifford algebra \cite{Fuchs_1997}, anticommutation relations,
\begin{equation}
  \{\gamma_j,\gamma_{j'}\}=2\delta_{jj'},\quad j,j'=1,2.
  \label{TS_operators_2}
\end{equation}
A finite value of the energy $\xi$ in Eq. (\ref{TS_Ham}) corresponds to a
finite overlap between the MBSs. The TS is assumed to be grounded.

The QD interacts with the metallic contacts and TS via tunneling mechanisms
modeled by the tunneling Hamiltonians $\hat{H}_\text{QD-C}$ and
$\hat{H}_\text{QD-TS}$, respectively.

The Hamiltonian for the tunneling between the QD and metallic contacts,
\begin{equation}
  \hat{H}_\text{QD-C}=\sum_{l=\{L,R\}}\mathcal{T}_l\sum_kc^\dagger_{lk}d+\text{H.c.},
  \label{Tunn_Ham_QD-C}
\end{equation}
uses the approximation
\begin{equation}
  \mathcal{T}_{lk}\approx\mathcal{T}_l
\end{equation}
for the tunneling matrix elements whose dependence on $k$ is sufficiently weak
in the energy range relevant for quantum transport measurements. The tunneling
interaction between the QD and metallic contacts appears via the energy scales
\begin{equation}
  \Gamma_{L,R}=\pi\nu_C|\mathcal{T}_{L,R}|^2.
\end{equation}
The sum of these energies,
\begin{equation}
  \Gamma=\Gamma_L+\Gamma_R,
\end{equation}
determines the escape rate of quasiparticles from the QD into the metallic
contacts. To simplify the model, in our numerical calculations we have assumed
that the QD is coupled symmetrically to the left and right contacts,
\begin{equation}
  \Gamma_L=\Gamma_R.
  \label{Symm_coupl}
\end{equation}

The Hamiltonian for the tunneling between the QD and TS,
\begin{equation}
  \hat{H}_\text{QD-TS}=\eta_1^\star d^\dagger\gamma_1+\eta_2^\star d^\dagger\gamma_2+\text{H.c.},
  \label{Tunn_Ham_QD-TS}
\end{equation}
includes in general two interactions, namely, between the QD and the two
Majorana modes, $\gamma_1$ and $\gamma_2$. The Majorana tunneling matrix
elements,
\begin{equation}
  \eta_{1,2}=|\eta_{1,2}|\,e^{i\phi_{1,2}},
  \label{M_tunn_matr_el}
\end{equation}
play the central role in the present research because they are responsible for
various interference effects in nanostructures with $|\eta_1|\neq 0$ and
$|\eta_2|\neq 0$. The Majorana interference arises when the tunneling phases
become unequal,
\begin{equation}
  \Delta\phi\equiv\phi_1-\phi_2\neq 0.
  \label{Delta_phi}
\end{equation}

For more clarity the above mathematical description of the system is
represented graphically in Fig. \ref{figure_1}. To be specific, in our
numerical calculations we assume a nanostructure similar to the one in
Ref. \cite{Deng_2018} where the QD is located asymmetrically with respect to
the two MBSs, {\it i.e.} much closer to one Majorana mode than to the
other. The numerical results presented in Sec. \ref{Results} have been
obtained for $|\eta_1|\gg|\eta_2|$ that is the Majorana mode $\gamma_1$ has
been chosen as the closest to the QD. We note, that even a very weak coupling
between $\gamma_2$ and the QD becomes crucial for fluctuations of
thermoelectric currents induced by both $V$ and $V_T$. Indeed, even when
$|\eta_2|$ is several orders of magnitude smaller than $|\eta_1|$, the
thermoelectric quantum noise at finite frequencies acquires a very strong
dependence on the Majorana tunneling phase difference $\Delta\phi$ as
demonstrated by the numerical results presented in Sec. \ref{Results}.
\section{Keldysh formalism for quantum fluctuations of nonequilibrium
  thermoelectric currents}\label{Keld}
If the chemical potentials and temperatures of the metallic contacts do not
vary in time, the nanoscopic system specified in Sec. \ref{Ham} will be
brought into a stationary nonequilibrium state produced by both the bias and
thermal voltages. To deal with stationary nonequilibrium we apply the
formalism of the Keldysh field integral \cite{Altland_2010} which is a
flexible mathematical tool developed on the basis of the original Keldysh
technique \cite{Keldysh_1965}. In particular, it allows one to consistently
generate current-current correlation functions which we need in this work to
analyze the thermoelectric quantum noise at finite frequencies in the presence
of Majorana interference effects. Although the Keldysh field integral is well
described in various textbooks \cite{Altland_2010}, below we briefly remind
some of its steps in the context of our problem.

Due to the fermionic nature of our system the Keldysh field integral is
performed over the Grassmann fields and their conjugate partners,
$(\psi,\bar{\psi})$, $(\phi_{lk},\bar{\phi}_{lk})$ and $(\zeta,\bar{\zeta})$,
appearing after the standard mapping \cite{Altland_2010} from the operators
$(d,d^\dagger)$, $(c_{lk},c_{lk}^\dagger)$ and $(\gamma_1,\gamma_2)$ to their
eigenvalues representing the states of the QD, contacts and TS,
respectively. Here and below the Grassmann conjugation (G.c.) is denoted with
an upper bar. The closed time contour may be parameterized by the real time
$t$ and the branch index $q$ labeling its forward ($q=+$) and backward ($q=-$)
branches. Then the Keldysh generating functional is written as a field
integral,
\begin{equation}
  \mathcal{Z}[\mathcal{J}_{lq}(t)]=\int\mathcal{D}[\bar{\Phi}_q(t),\Phi_q(t)]e^{\frac{i}{\hbar}S_\text{K}[\mathcal{J}_{lq}(t)]},
  \label{Gen_functional}
\end{equation}
over the corresponding Grassmann fields and their conjugate partners,
\begin{equation}
  \begin{split}
    &\{\Phi_q(t);\,\bar{\Phi}_q(t)\}\equiv\\
    \equiv&\{\psi_q(t),\,\phi_{lkq}(t),\,\zeta_q(t);\,\bar{\psi}_q(t),\,\bar{\phi}_{lkq}(t),\,\bar{\zeta}_q(t)\},
  \end{split}
  \label{Grassmann_fields}
\end{equation}
defined on the real time axis $t$. The Keldysh action $S_K$,
\begin{equation}
  \begin{split}
    &S_\text{K}[\mathcal{J}_{lq}(t)]=S_\text{QD}[\bar{\psi}_q(t),\psi_q(t)]+\\
    &+S_\text{C}[\bar{\phi}_{lkq}(t),\phi_{lkq}(t)]+S_\text{TS}[\bar{\zeta}_q(t),\zeta_q(t)]+\\
    &+S_\text{QD-C}[\bar{\psi}_q(t),\bar{\phi}_{lkq}(t);\psi_q(t),\phi_{lkq}(t)]+\\
    &+S_\text{QD-TS}[\bar{\psi}_q(t),\bar{\zeta}_q(t);\psi_q(t),\zeta_q(t)]+S_\text{SRC}[\mathcal{J}_{lq}(t)],
  \end{split}
  \label{Keldysh_action}
\end{equation}
is a functional of the source field $\mathcal{J}_{lq}(t)$ introduced via the
source action $S_\text{SRC}$ to generate current-current correlators. Note,
that the generating functional is properly normalized,
\begin{equation}
  \mathcal{Z}[\mathcal{J}_{lq}(t)=0]=1.
  \label{Gen_functional_norm}
\end{equation}
After the Keldysh rotation \cite{Altland_2010} the Grassmann fields on the
forward and backward branches are combined into the retarded and advanced
components. This transformation brings the actions of the QD, contacts and TS,
denoted as $S_\text{QD}$, $S_\text{C}$ and $S_\text{TS}$, into the
conventional $2\times 2$ matrix form \cite{Altland_2010}. The action
describing tunneling processes between the QD and contacts mixes the Grassmann
fields $\{\bar{\psi}_q(t),\psi_q(t)\}$ and
$\{\bar{\phi}_{lkq}(t),\phi_{lkq}(t)\}$:
\begin{equation}
  \begin{split}
    &S_\text{QD-C}[\bar{\psi}_q(t),\bar{\phi}_{lkq}(t);\psi_q(t),\phi_{lkq}(t)]=\\
    &=-\int_{-\infty}^\infty dt\sum_{l=\{L,R\}}\sum_{k,q}\bigl[\mathcal{T}_lq\bar{\phi}_{lkq}(t)\psi_q(t)+\text{G.c.}\bigr].
  \end{split}
  \label{Tunn_act_QD-C}
\end{equation}
The action taking into account the tunneling interaction between the QD and TS
mixes the Grassmann fields $\{\bar{\psi}_q(t),\psi_q(t)\}$ with
$\{\bar{\zeta}_q(t),\zeta_q(t)\}$,
\begin{equation}
  \begin{split}
    &S_\text{QD-TS}[\bar{\psi}_q(t),\bar{\zeta}_q(t);\psi_q(t),\zeta_q(t)]=\\
    &=-\int_{-\infty}^\infty dt\sum_qq\bigl\{\eta_1^\star \bigl[\bar{\psi}_q(t)\zeta_q(t)+\bar{\psi}_q(t)\bar{\zeta}_q(t)\bigr]+\\
    &+i\eta_2^\star \bigl[\bar{\psi}_q(t)\zeta_q(t)+\bar{\psi}_{-q}(t)\bar{\zeta}_{-q}(t)\bigr]+\text{G.c.}\bigr\},
  \end{split}
  \label{Tunn_act_QD-TS}
\end{equation}
and involves the Majorana tunneling phases $\phi_1$ and $\phi_2$ through the
matrix elements $\eta_1$ and $\eta_2$. This tunneling action is of particular
importance because various Majorana interference effects in physical
observables result eventually from Eq. (\ref{Tunn_act_QD-TS}).

As mentioned above, the second quantized operators of the QD, contacts and TS
map to the corresponding Grassmann-valued fields. As a result, the operators
of physical observables map to the corresponding functions of these fields. In
particular, the current operator in contact $l=\{L,R\}$,
\begin{equation}
  \hat{I}_l=\frac{ie}{\hbar}\sum_k\bigl(\mathcal{T}_lc_{lk}^\dagger d-\text{H.c.}\bigr),
  \label{Current_op}
\end{equation}
maps to the Grassmann-valued field
\begin{equation}
  I_{lq}(t)=\frac{ie}{\hbar}\sum_k\bigl[\mathcal{T}_l\bar{\phi}_{lkq}(t)\psi_q(t)-\text{G.c.}\bigr].
  \label{Current_Grassmann}
\end{equation}
Using this field one may define the source action $S_\text{SRC}$,
\begin{equation}
  S_\text{SRC}[\mathcal{J}_{lq}(t)]=-\int_{-\infty}^\infty dt\sum_{l=\{L,R\}}\sum_q\mathcal{J}_{lq}(t)I_{lq}(t),
  \label{Source_act}
\end{equation}
to generate current-current correlators,
\begin{equation}
  \langle I_{lq}(t)I_{l'q'}(t')\rangle_0=(i\hbar)^2\frac{\delta^2\mathcal{Z}[\mathcal{J}_{lq}(t)]}{\delta\mathcal{J}_{lq}(t)\delta\mathcal{J}_{l'q'}(t')}\biggl|_{\mathcal{J}_{lq}(t)=0},
  \label{Curr-Curr_corr}
\end{equation}
where the averaging $\langle\ldots\rangle_0$ is defined with respect to the
Keldysh action
\begin{equation}
  S_\text{K}^{(0)}\equiv S_\text{K}[\mathcal{J}_{lq}(t)=0],
  \label{S_K0}
\end{equation}
that is
\begin{equation}
  \begin{split}
    &\bigl\langle I_{l_1q_1}(t_1)\cdots I_{l_nq_n}(t_n)\bigr\rangle_0\equiv\\
    &\equiv\int\mathcal{D}[\bar{\Phi}_q(t),\Phi_q(t)]e^{\frac{i}{\hbar}S_\text{K}^{(0)}}I_{l_1q_1}(t_1)\cdots I_{l_nq_n}(t_n).
  \end{split}
  \label{Average_SK0}
\end{equation}
Note, that in stationary nonequilibrium the average in Eq. (\ref{Average_SK0})
depends on the temporal arguments only through the differences $t_1-t_k$,
$k=2,\ldots ,n$. To obtain the mean current in contact $l=\{L,R\}$ one may
choose any branch $q$ in Eq. (\ref{Current_Grassmann}) since after averaging
the dependence on $q$ disappears,
\begin{equation}
  I_l(V,V_T,\Delta\phi)=\bigl\langle I_{lq}(t)\bigr\rangle_0=i\hbar\frac{\delta\mathcal{Z}[\mathcal{J}_{lq}(t)]}{\delta\mathcal{J}_{lq}(t)}\biggl|_{\mathcal{J}_{lq}(t)=0}.
  \label{Mean_current}
\end{equation}

In general one may consider various current-current correlators either within
a given contact or between different contacts. To be specific, in what follows
we will investigate fluctuations of the thermoelectric current in the left
metallic contact and characterize these fluctuations by the quantum noise
specified via the greater current-current correlator:
\begin{equation}
  S^>(t-t',V,V_T,\Delta\phi)\equiv\bigl\langle\delta I_{L-}(t)\delta I_{L+}(t')\bigr\rangle_0,
  \label{QN_time_domain}
\end{equation}
where
\begin{equation}
  \delta I_{Lq}(t)\equiv I_{Lq}(t)-I_L(V,V_T,\Delta\phi)
  \label{Curr_dev_Grassman}
\end{equation}
is the Grassmann-valued field corresponding to the current deviations from the
mean value. The calculation of the average $\langle\ldots\rangle_0$ in
Eq. (\ref{QN_time_domain}) is performed using conventional Wick's theorem as
in Ref. \cite{Smirnov_2018} where one can find more technical details.

In the frequency domain,
\begin{equation}
  S^>(\omega,V,V_T,\Delta\phi)=\int_{-\infty}^\infty dt\,e^{i\omega t}S^>(t,V,V_T,\Delta\phi),
  \label{QN_freq_domain}
\end{equation}
the quantum noise provides a unique way to explore Majorana interference
effects via a thermoelectric fluctuation response at finite frequencies in
both the absorption,
\begin{equation}
  S^\text{ab}(\omega,V,V_T,\Delta\phi)=S^>(\omega>0,V,V_T,\Delta\phi),
  \label{QN_ab}
\end{equation}
and emission,
\begin{equation}
  S^\text{em}(\omega,V,V_T,\Delta\phi)=S^>(\omega<0,V,V_T,\Delta\phi),
  \label{QN_em}
\end{equation}
photon spectra. To be specific, below we address the fluctuation response
encoded at finite frequencies in the differential thermoelectric quantum noise
defined as the derivative with respect to the thermal voltage,
$\partial S^>(\omega,V,V_T,\Delta\phi)/\partial V_T$, which might be of
experimental interest similarly to the differential quantum noise
\cite{Basset_2012} defined as the derivative with respect to the bias
voltage.
\section{Finite frequency behavior of the differential thermoelectric quantum
  noise in various regimes}\label{Results}
To obtain the results presented in this section we have implemented two
numerical approaches. In the first approach we have analytically calculated
all the necessary derivatives over the thermal voltage $V_T$ since it enters
only through the Fermi-Dirac distribution of the left metallic contact as may
be seen in Eqs. (\ref{F-D}), (\ref{Temp_L_R}) and (\ref{Therm_volt}). After
that numerical integrations have been used to obtain
$\partial S^>(\omega,V,V_T,\Delta\phi)/\partial V_T$. In the second approach,
numerical integrations are first performed to obtain
$S^>(\omega,V,V_T,\Delta\phi)$. After that we have used finite differences to
numerically calculate the derivative
$\partial S^>(\omega,V,V_T,\Delta\phi)/\partial V_T$. In different physical
regimes the difference between the computational times of these two approaches
may be significant and we have always used the faster one. It is important to
note that various outputs obtained by the two approaches have been compared
and found to be identical. This provides a good verification for the numerical
results presented below.

To study the universal Majorana behavior, that is the one independent of
$\epsilon_d$, the numerical calculations have been restricted to the regime in
which one of the Majorana tunneling amplitudes, $|\eta_1|$ or $|\eta_2|$, is
the largest energy scale of the problem. Since in Sec. \ref{Ham} we have
assumed $|\eta_1|\gg|\eta_2|$, we have chosen $|\eta_1|$ as the largest energy
scale:
\begin{equation}
  |\eta_1|>\text{max}\bigl\{|\epsilon_d|,k_\text{B}T,|\eta_2|,\xi,\Gamma,eV_T,|eV|\bigr\}.
  \label{Majorana_regime}
\end{equation}
In the universal Majorana regime specified by Eq. (\ref{Majorana_regime})
physical observables do not depend on $\epsilon_d$ and thus its value may be
chosen arbitrarily within this regime. The numerical results presented below
have been obtained for the case $\epsilon_d>0$ but they remain the same also
for the case $\epsilon_d<0$. In possible experiments, however, the magnetic
field inducing the topological phase with emerging MBSs may strongly but still
incompletely suppress the Kondo correlations in QDs with $\epsilon_d<0$ and,
as a result, the Kondo universality
\cite{Hewson_1997,Smirnov_2011a,Niklas_2016} may superimpose on the Majorana
one. Thus, to avoid any possible interplay
\cite{Cheng_2014,Weymann_2020,Majek_2022,Wojcik_2024} between the Kondo effect
and MBSs, experiments with $\epsilon_d>0$ would be preferable to observe
phenomena driven by exclusively MBSs.
\begin{figure}
  \includegraphics[width=8.0 cm]{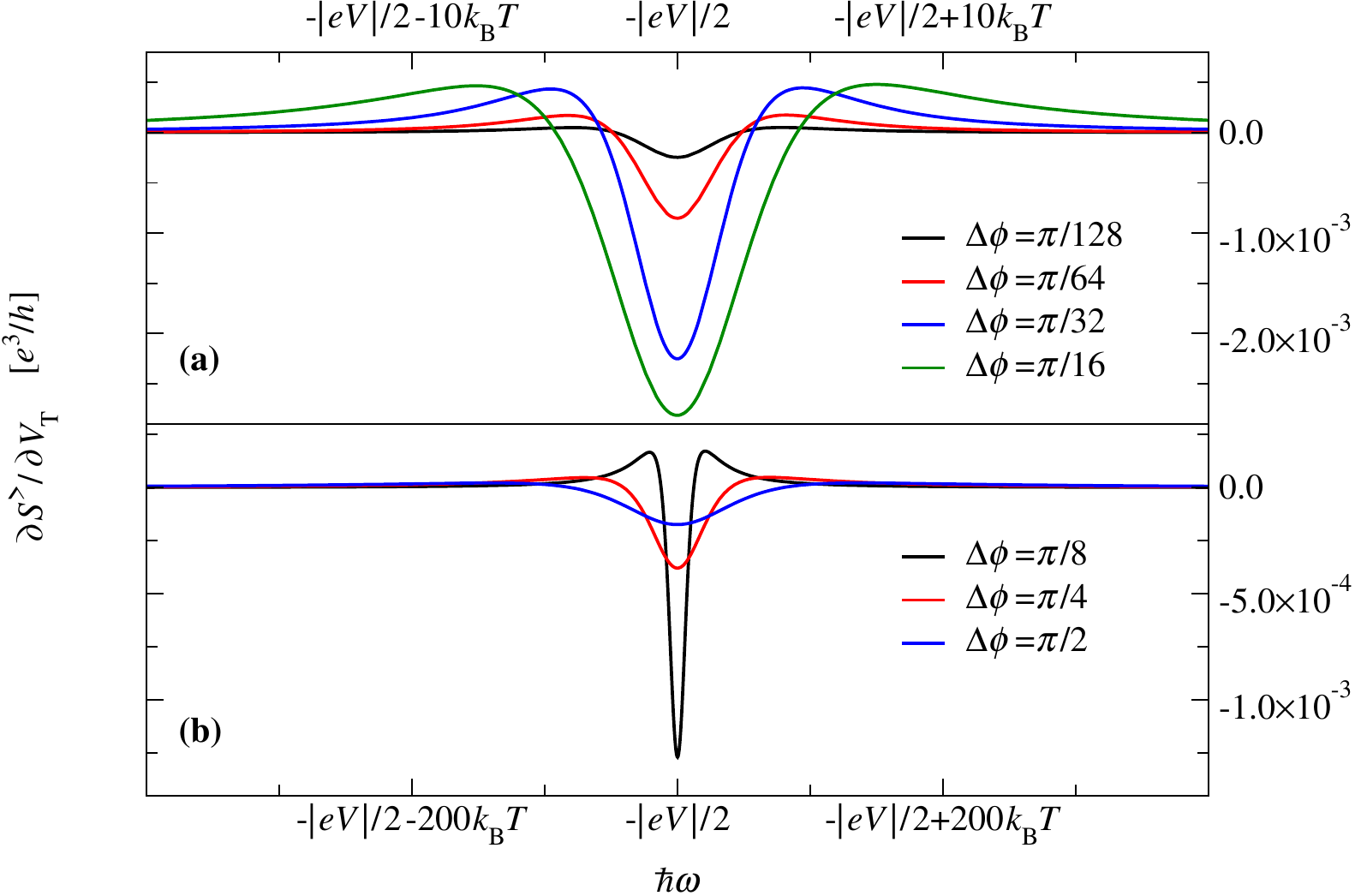}
  \caption{\label{figure_2} Frequency dependence of the differential
    thermoelectric quantum noise,
    $\partial S^>(\omega,V,V_T,\Delta\phi)/\partial V_T$, in a vicinity of the
    frequency $\hbar\omega=-|eV|/2$ (emission noise, domain $\omega<0$) for
    small bias voltages, $|eV|\ll\Gamma$, and for various values of the Majorana
    tunneling phase difference $\Delta\phi$. Specifically, for all the curves
    $|eV|/\Gamma=10^{-2}$ and, in Panel ({\bf a}), $\Delta\phi=\pi/128$
    (black), $\Delta\phi=\pi/64$ (red), $\Delta\phi=\pi/32$ (blue),
    $\Delta\phi=\pi/16$ (green), in Panel ({\bf b}), $\Delta\phi=\pi/8$
    (black), $\Delta\phi=\pi/4$ (red), $\Delta\phi=\pi/2$ (blue). The values
    of the other parameters: $\epsilon_d/\Gamma=10$,
    $k_\text{B}T/\Gamma=10^{-7}$, $eV_T/\Gamma=10^{-9}$,
    $|\eta_1|/\Gamma=10^3$, $|\eta_2|/\Gamma=10^{-3}$, $\xi/\Gamma=10^{-4}$.}
\end{figure}

In Fig. \ref{figure_2} we show the differential thermoelectric quantum noise
as a function of the frequency in the domain $\omega<0$ (emission
noise) for small bias voltages, $|eV|\ll\Gamma$. Here, in addition to the
resonance located at $\hbar\omega=-|eV|$, which has been discussed previously
\cite{Smirnov_2019a}, there arises qualitatively new behavior around the
frequency $\hbar\omega=-|eV|/2$. When the energy scales are well separated,
that is when $k_\text{B}T$ is several orders of magnitude less than $|eV|$ (in
the regime $eV_T\ll k_\text{B}T\ll|eV|$) or $eV_T$ is several orders of
magnitude less than $|eV|$ (in the regime $k_\text{B}T\ll eV_T\ll|eV|$), there
develops an antiresonance with its minimum located at
$\hbar\omega=-|eV|/2$. For small values of the Majorana tunneling phase
difference $\Delta\phi$ the width of this antiresonance is determined either
by $k_\text{B}T$, as in the present regime $eV_T\ll k_\text{B}T\ll|eV|$ (black
and red curves in Fig. \ref{figure_2}(a)), or by $eV_T$, in the regime
$k_\text{B}T\ll eV_T\ll|eV|$. It is important to note that this antiresonance
appears when $|\eta_2|\neq 0$ and $\Delta\phi\neq 0$ and does not arise when
the QD is effectively coupled to a single MBS, that is for $\Delta\phi=0$ or
$|\eta_2|=0$. Note also, that a difference in several orders of magnitude
between the corresponding energy scales mentioned above is crucial for its
appearance. Indeed, if this difference was only one order of magnitude or
less, the antiresonance would be strongly suppressed because the resonances
located at $\hbar\omega=0$ and $\hbar\omega=-|eV|$ (see
Ref. \cite{Smirnov_2019a}) would overlap and fully wash out the antiresonance
at $\hbar\omega=-|eV|/2$. However, when the energy scales are separated by
several orders of magnitude, as in the present case, the resonances located at
$\hbar\omega=0$ and $\hbar\omega=-|eV|$ are far away from each other and there
emerges the antiresonance at $\hbar\omega=-|eV|/2$ shown in
Fig. \ref{figure_2}. The minimum of this antiresonance is located exactly at
$\hbar\omega=-|eV|/2$. Both its amplitude and width depend on
$\Delta\phi$. The amplitude of the antiresonance first grows with
$\Delta\phi$, reaches a maximal value at an intermediate value of $\Delta\phi$
(see Fig. \ref{figure_2}(a)), and after that decreases to a small but finite
value when the Majorana phase difference goes to $\Delta\phi=\pi/2$ (see
Fig. \ref{figure_2}(b)). In contrast, as can be seen from
Figs. \ref{figure_2}(a) and (b), the width of the emission antiresonance
always grows with $\Delta\phi$ from a value of the order of $k_\text{B}T$, at
small values of $\Delta\phi$, to very large values, about $10^2k_\text{B}T$,
at $\Delta\phi=\pi/2$. The appearance of this antiresonance qualitatively
changes the behavior of the differential thermoelectric quantum noise. Indeed,
we find that in the absorption spectra ($\omega>0$) there does not arise an
antiresonance at $\hbar\omega=|eV|/2$ which could have been the absorption
partner of the emission antiresonance. As a result,
$\partial S^>(\omega,V,V_T,\Delta\phi)/\partial V_T$ loses its symmetry in
$\omega$ and the emission and absorption spectra are in general not related to
each other. This is in contrast to the symmetric differential thermoelectric
quantum noise observed in Ref. \cite{Smirnov_2019a} for the case
$|\eta_2|=0$.
\begin{figure}
  \includegraphics[width=8.0 cm]{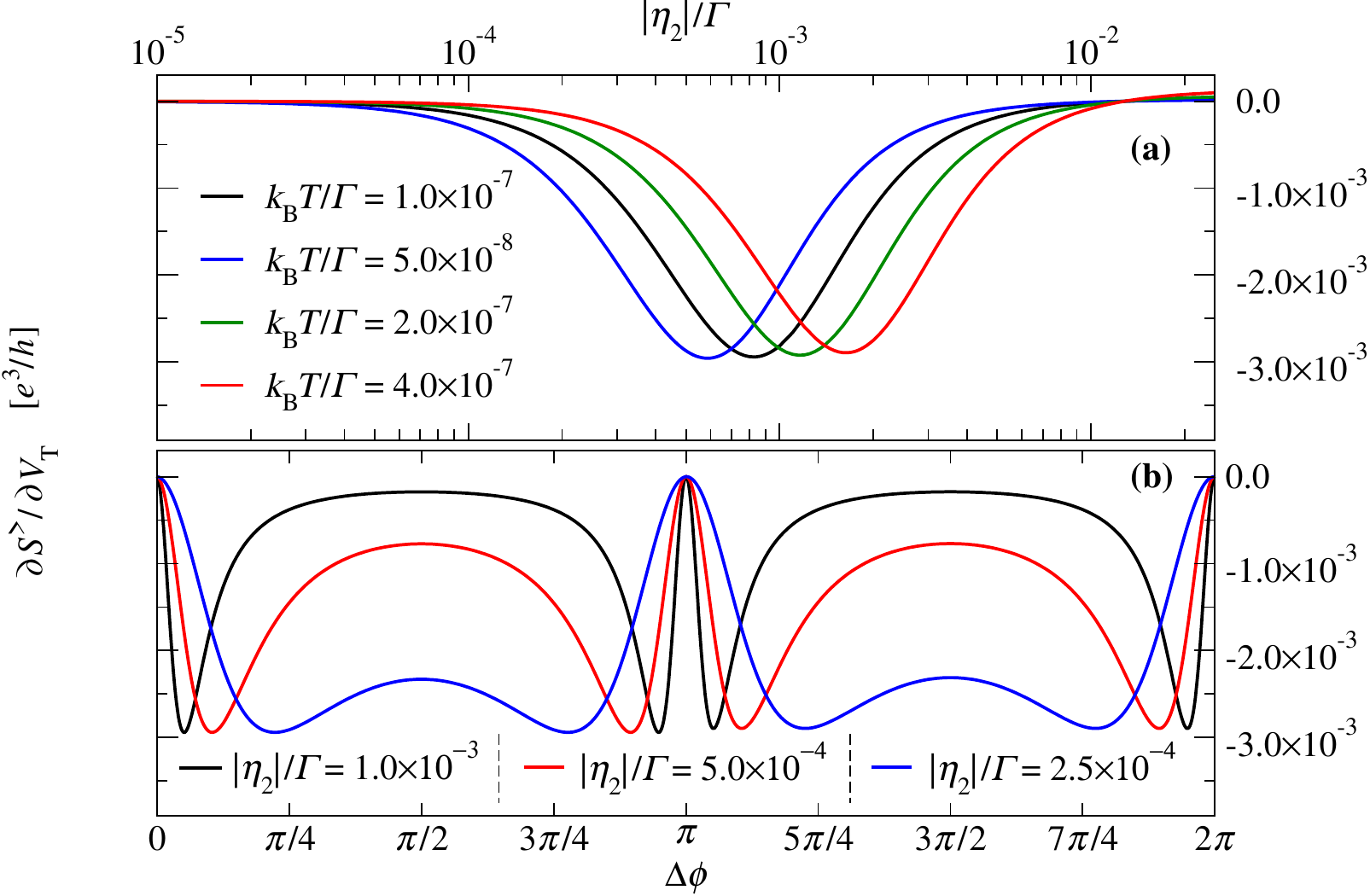}
  \caption{\label{figure_3} Differential thermoelectric quantum noise,
    $\partial S^>(\omega,V,V_T,\Delta\phi)/\partial V_T$, at
    $\hbar\omega=-|eV|/2$. Panel ({\bf a}): As a function of $|\eta_2|$ for
    $\Delta\phi=\pi/16$ and four values of the temperature
    $k_\text{B}T$. Specifically, $k_\text{B}T/\Gamma=10^{-7}$ (black),
    $k_\text{B}T/\Gamma=5.0\times 10^{-8}$ (blue),
    $k_\text{B}T/\Gamma=2.0\times 10^{-7}$ (green),
    $k_\text{B}T/\Gamma=4.0\times 10^{-7}$ (red). Panel ({\bf b}): As a
    function of $\Delta\phi$ for $k_\text{B}T/\Gamma=10^{-7}$ and three values
    of $|\eta_2|$. Specifically, $|\eta_2|/\Gamma=10^{-3}$ (black),
    $|\eta_2|/\Gamma=5.0\times 10^{-4}$ (red),
    $|\eta_2|/\Gamma=2.5\times 10^{-4}$ (blue). The other parameters are
    the same as in Fig. \ref{figure_2}.}
\end{figure}

To further analyze the behavior of the emission antiresonance we have
calculated the differential thermoelectric quantum noise at
$\hbar\omega=-|eV|/2$ as a function of $|\eta_2|$ and $\Delta\phi$ for
different temperatures as shown in Fig. \ref{figure_3}. From the dependence on
the Majorana tunneling amplitude $|\eta_2|$, demonstrated in
Fig. \ref{figure_3}(a), one can see that the curves for different temperatures
are characterized by a minimum where the amplitude of the emission
antiresonance is maximal. The minima of all the curves are almost the
same. This means that the maximal amplitude of the emission antiresonance
weakly depends on the temperature whose effect is just to shift it to another
value of $|\eta_2|$. To physically interpret such a shift, it is important to
note that thermal fluctuations represent a mechanism suppressing the Majorana
interference. Therefore they will compete against the fluctuations induced by
the interfering MBSs when $|\eta_2|\neq 0$ and $\Delta\phi\neq 0$. Indeed, as
demonstrated in Fig. \ref{figure_3}(a), when the temperature grows, thermal
fluctuations push out the emission antiresonance from weak to strong Majorana
interference, that is from smaller to larger values of $|\eta_2|$. Thus, in
order to restore this antiresonance after it has been destroyed by thermal
fluctuations, one has to increase $|\eta_2|$ to enhance the fluctuations
induced by the Majorana interference. However, the increase of $|\eta_2|$
should not be too large since very strong interference becomes destructive
and, as can be seen in Fig. \ref{figure_3}(a), ruins the antiresonance. Note
also, that larger values of $|\eta_2|$ shrink the range of $\Delta\phi$ where
the amplitude of the emission antiresonance is large enough to be observed in
experiments. This may be seen from the black curve in
Fig. \ref{figure_3}(b). At the same time the blue curve in
Fig. \ref{figure_3}(b) shows that for smaller values of $|\eta_2|$ the
amplitude of the emission antiresonance becomes larger in a wide range of
$\Delta\phi$. Since for $\Delta\phi=0,\pi$ the QD is effectively coupled to a
single MBS, the emission antiresonance is absent at $\Delta\phi=0,\pi$ for any
finite value of $|\eta_2|$ and its amplitude is strongly suppressed in
vicinities of these points. In the limit $|\eta_2|\rightarrow 0$ these
vicinities will widen indicating the disappearance of the Majorana
interference for any finite value of $\Delta\phi$.
\begin{figure}
  \includegraphics[width=8.0 cm]{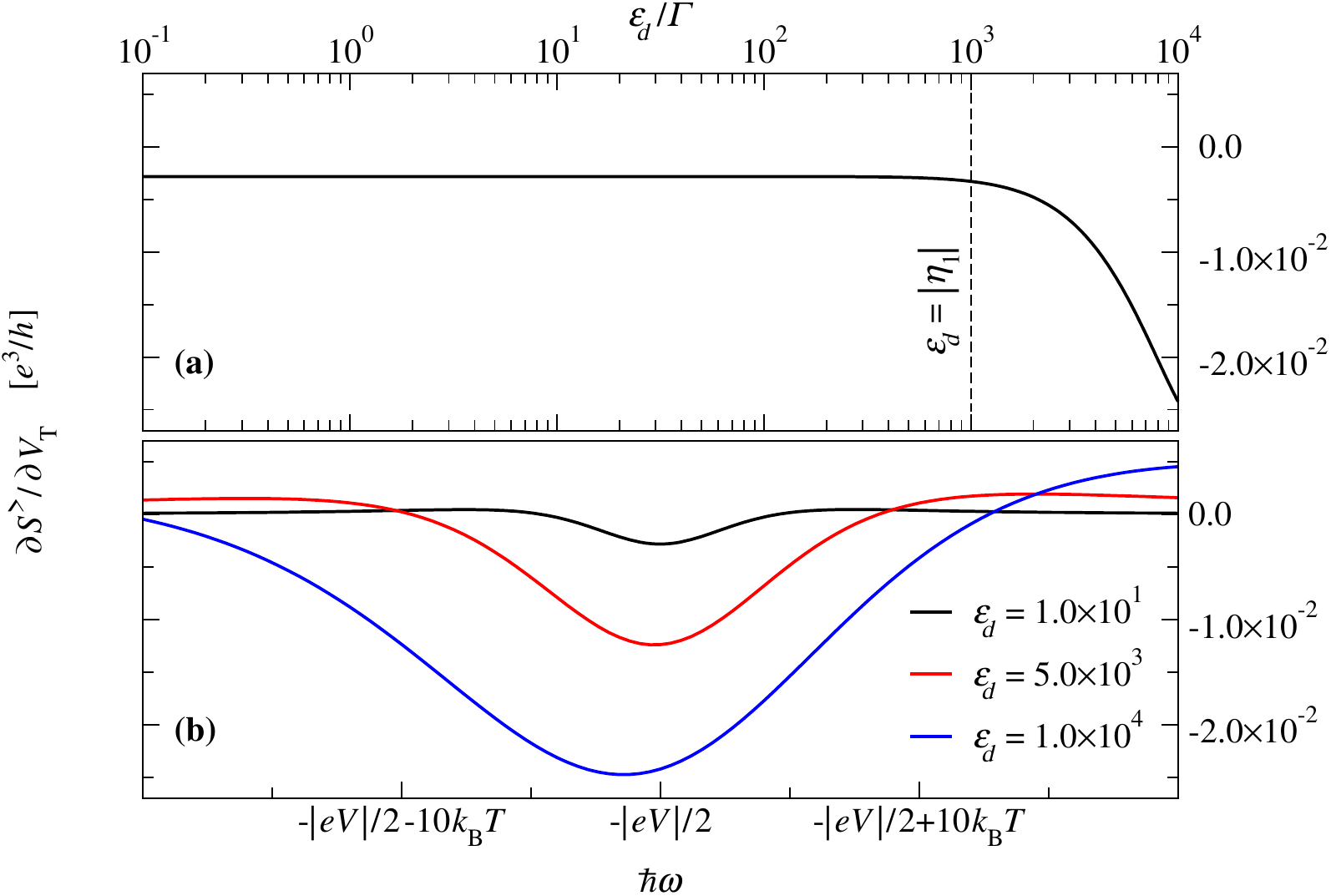}
  \caption{\label{figure_4} Differential thermoelectric quantum noise,
    $\partial S^>(\omega,V,V_T,\Delta\phi)/\partial V_T$, in the emission
    domain ($\omega<0$). Panel ({\bf a}): As a function of the gate voltage
    $\epsilon_d$ at the frequency $\hbar\omega=-|eV|/2$. Panel ({\bf b}): As a
    function of the frequency $\hbar\omega$ in a vicinity of
    $\hbar\omega=-|eV|/2$ for various values of the gate voltage
    $\epsilon_d$. Specifically, $\epsilon_d/\Gamma=10$ (black),
    $\epsilon_d/\Gamma=5.0\times 10^3$ (red), $\epsilon_d/\Gamma=10^4$
    (blue). In both panels $\Delta\phi=\pi/16$. The other parameters are the
    same as in Fig. \ref{figure_2}.}
\end{figure}

Let us look how the results shown above change when one moves outside the
universal Majorana regime (see Eq. (\ref{Majorana_regime})) as may happen
during possible experiments. As has been mentioned above, within the universal
Majorana regime the differential thermoelectric quantum noise is independent
of the gate voltage. This may be seen from Fig. \ref{figure_4}(a), which shows
$\partial S^>(\omega,V,V_T,\Delta\phi)/\partial V_T$ as a function of the gate 
voltage $\epsilon_d$ at the minimum of the emission antiresonance, that is at
$\hbar\omega=-|eV|/2$. As can be seen, the minimum of the emission
antiresonance is independent of $\epsilon_d$ for gate voltages satisfying the
condition in Eq. (\ref{Majorana_regime}), $\epsilon_d<|\eta_1|$. Weak
deviations start to appear when $\epsilon_d\sim|\eta_1|$. For larger gate
voltages, $\epsilon_d>|\eta_1|$, the minimum of the emission antiresonance
acquires a strong dependence on $\epsilon_d$. The same is also true at other
frequencies so that the whole antiresonance remains unchanged when
$\epsilon_d$ is varied within its upper bound, $\epsilon_d<|\eta_1|$. It turns
out that for larger gate voltages, $\epsilon_d>|\eta_1|$, the emission
antiresonance is still present in a vicinity of the frequency
$\hbar\omega=-|eV|/2$. However, as demonstrated by Fig. \ref{figure_4}(b),
when $\epsilon_d$ starts to exceed $|\eta_1|$, the emission antiresonance
acquires two qualitative changes. First, its minimum shifts from
$\hbar\omega=-|eV|/2$. Second, it becomes asymmetric with respect to its
minimum. Both the shift of the minimum and asymmetry get larger and larger as
$\epsilon_d$ more and more exceeds $|\eta_1|$ as can be seen from the red and
blue curves in Fig. \ref{figure_4}(b). Additionally, the antiresonance
exhibits quantitative changes, namely, its amplitude and width significantly
increase which might facilitate its experimental detection.
\begin{figure}
  \includegraphics[width=8.0 cm]{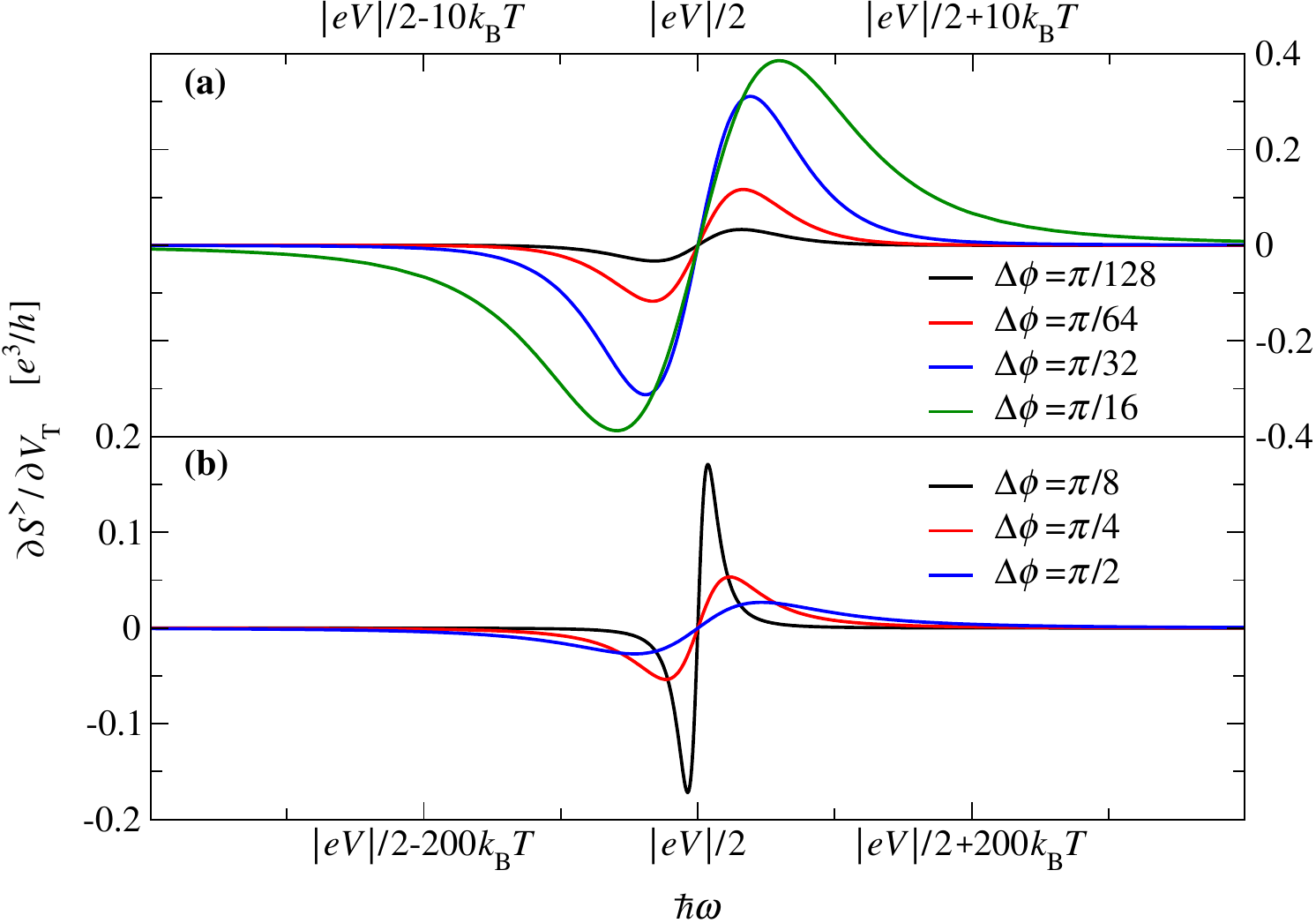}
  \caption{\label{figure_5} Frequency dependence of the differential
    thermoelectric quantum noise,
    $\partial S^>(\omega,V,V_T,\Delta\phi)/\partial V_T$, in a vicinity of the
    frequency $\hbar\omega=|eV|/2$ (absorption noise, domain $\omega>0$) for
    small bias voltages, $|eV|\ll\Gamma$, and for various values of the Majorana
    tunneling phase difference $\Delta\phi$. Specifically, for all the curves
    $|eV|/\Gamma=10^{-2}$ and, in Panel ({\bf a}), $\Delta\phi=\pi/128$ (black),
    $\Delta\phi=\pi/64$ (red), $\Delta\phi=\pi/32$ (blue), $\Delta\phi=\pi/16$
    (green), in Panel ({\bf b}), $\Delta\phi=\pi/8$ (black), $\Delta\phi=\pi/4$
    (red), $\Delta\phi=\pi/2$ (blue). The other parameters have the same values
    as in Fig. \ref{figure_2}.}
\end{figure}

The behavior of the absorption noise ($\omega>0$) in the vicinity of the
frequency $\hbar\omega=|eV|/2$ turns out to be qualitatively different from
the emission noise ($\omega<0$) in the vicinity of the frequency
$\hbar\omega=-|eV|/2$. When the QD is effectively coupled to a single MBS,
that is when $|\eta_2|=0$ or $\Delta\phi=0$, the differential thermoelectric
quantum noise is strongly suppressed in the vicinity of the frequency
$\hbar\omega=|eV|/2$ and the absorption spectra are characterized by the
resonance at $\hbar\omega=|eV|$ as discussed in
Ref. \cite{Smirnov_2019a}. However, for finite values of $|\eta_2|$ the
differential thermoelectric quantum noise in the vicinity of
$\hbar\omega=|eV|/2$ becomes finite when one increases the Majorana tunneling
phase difference $\Delta\phi$ from zero to finite values. Indeed,
Fig. \ref{figure_5}(a) shows that for small values of $\Delta\phi$ there
develops an antiresonance-resonance pair whose center is located exactly at
$\hbar\omega=|eV|/2$. To be more specific,
$\partial S^>(\omega,V,V_T,\Delta\phi)/\partial V_T=0$ at the frequency
$\hbar\omega=|eV|/2$ and reaches a negative minimum and positive maximum with
equal amplitudes at some frequencies $\hbar\omega_{min}<|eV|/2$ and
$\hbar\omega_{max}>|eV|/2$, respectively. The frequencies $\hbar\omega_{min}$
and $\hbar\omega_{max}$ are equidistant from the center of the
antiresonance-resonance pair, that is from the frequency
$\hbar\omega=|eV|/2$. For small values of $\Delta\phi$ the amplitudes of the
antiresonance and resonance are also small (see the black and red curves with
$\Delta\phi=\pi/128$ and $\Delta\phi=\pi/64$) and their widths are of the
order of the temperature $k_\text{B}T$. For larger values of $\Delta\phi$ (see
the green curve with $\Delta\phi=\pi/16$) the amplitudes of the antiresonance
and resonance quickly grow. Their widths increase almost one order of
magnitude so that their extension in the frequency domain is almost
$10k_\text{B}T$. However, after their increase the amplitudes of the
antiresonance and resonance start to decrease at some value of $\Delta\phi$
and are significantly reduced when the Majorana tunneling phase difference
approaches the value $\Delta\phi=\pi/2$ as demonstrated in
Fig. \ref{figure_5}(b). In contrast, their widths continue to grow and their
extension in the frequency domain becomes very large, about $10^2k_\text{B}T$
(see the blue curve with $\Delta\phi=\pi/2$). Similar to the emission
antiresonance the absorption antiresonance-resonance pair is of pure Majorana
interference nature because it is strongly suppressed when the QD is
effectively coupled to a single MBS that is when $|\eta_2|=0$ or
$\Delta\phi=0$ and, as mentioned above, the emission and absorption spectra
are symmetrically related \cite{Smirnov_2019a}. However, when $|\eta_2|\neq 0$
and $\Delta\phi\neq 0$, the behavior of
$\partial S^>(\omega,V,V_T,\Delta\phi)/\partial V_T$ in the domains $\omega<0$
and $\omega>0$ is qualitatively different at small bias voltages,
$|eV|\ll\Gamma$, and the emission and absorption spectra are not related to
each other. This situation changes for large bias voltages, $|eV|\gg\Gamma$,
as will be shown below.

\begin{figure}
  \includegraphics[width=8.0 cm]{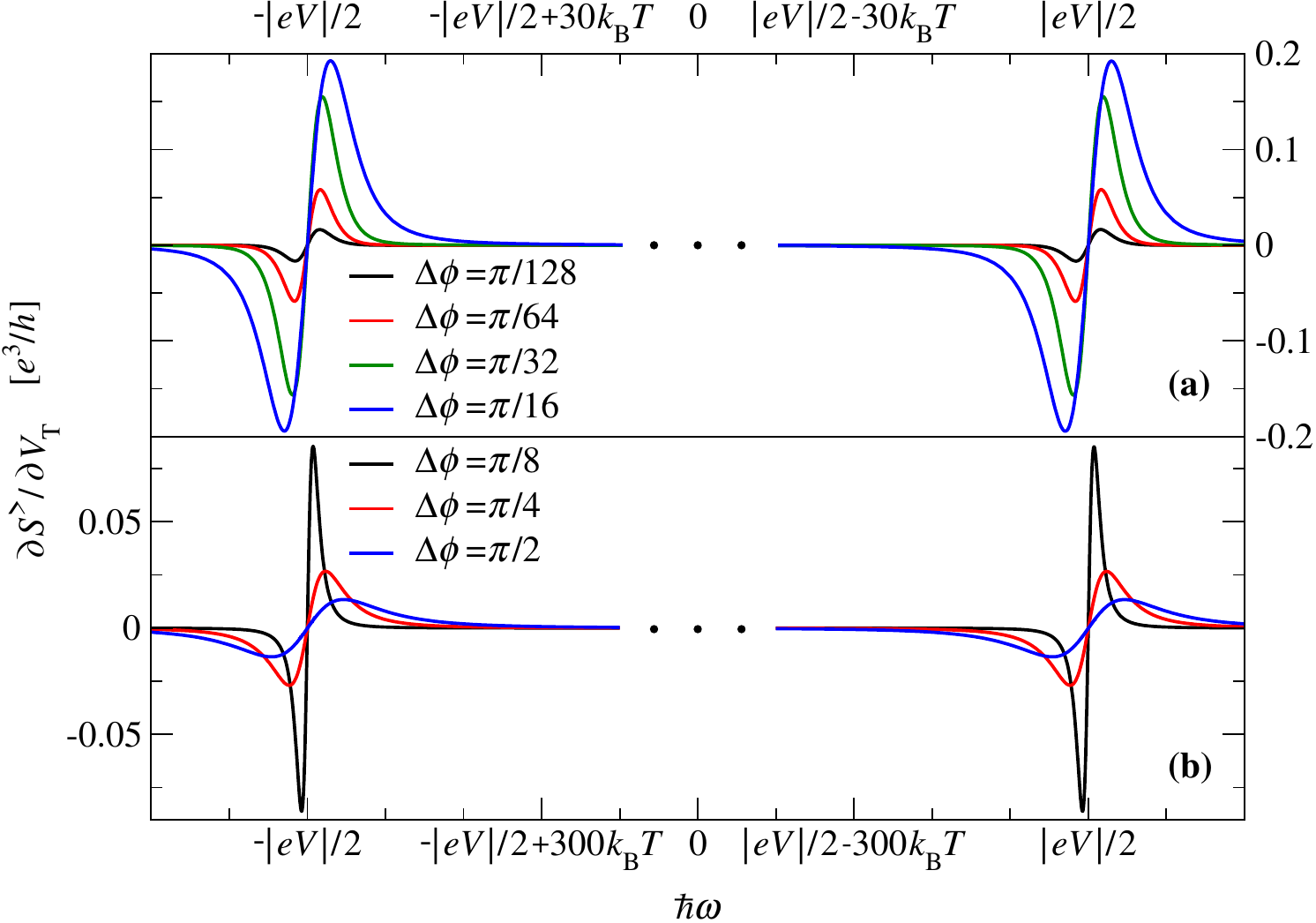}
  \caption{\label{figure_6} Frequency dependence of the differential
    thermoelectric quantum noise,
    $\partial S^>(\omega,V,V_T,\Delta\phi)/\partial V_T$, in the vicinity of the
    two frequencies, $\hbar\omega=-|eV|/2$ (emission noise, domain $\omega<0$)
    and $\hbar\omega=|eV|/2$ (absorption noise, domain $\omega>0$), for large
    bias voltages, $|eV|\gg\Gamma$, and for various values of the Majorana
    tunneling phase difference $\Delta\phi$. Specifically, for all the curves
    $|eV|/\Gamma=10^2$ and, in Panel ({\bf a}), $\Delta\phi=\pi/128$ (black),
    $\Delta\phi=\pi/64$ (red), $\Delta\phi=\pi/32$ (green), $\Delta\phi=\pi/16$
    (blue), in Panel ({\bf b}), $\Delta\phi=\pi/8$ (black), $\Delta\phi=\pi/4$
    (red), $\Delta\phi=\pi/2$ (blue). The other parameters have the same values
    as in Fig. \ref{figure_2}.}
\end{figure}

At large bias voltages, $|eV|\gg\Gamma$, it turns out that the differential
thermoelectric quantum noise is suppressed at all frequencies except for
vicinities of the frequencies $\hbar\omega=\pm|eV|/2$. In particular, when the
bias voltage is increased to large values, $|eV|\gg\Gamma$, the absorption
antiresonance-resonance pair discussed above in the regime $|eV|\ll\Gamma$
(see Fig. \ref{figure_5}) is retained at $\hbar\omega=|eV|/2$ but, as we find
numerically, the amplitudes of the antiresonance and resonance are reduced
exactly twice in the regime $|eV|\gg\Gamma$ as can be seen in
Fig. \ref{figure_6}. This happens for all values of the Majorana tunneling
phase difference $\Delta\phi$. The behavior of the widths of the antiresonance
and resonance in the regime $|eV|\gg\Gamma$ is the same as in the regime
$|eV|\ll\Gamma$. Specifically, for small values of $\Delta\phi$ the widths are
of the order of the temperature $k_BT$ (black and red curves in
Fig. \ref{figure_6}(a)), for larger values of $\Delta\phi$ the antiresonance
and resonance widen up to $10k_\text{B}T$ (blue curve in
Fig. \ref{figure_6}(a)) whereas for $\Delta\phi=\pi/2$ (blue curve in
Fig. \ref{figure_6}(b)) their widths are maximal, about $10^2k_\text{B}T$ in
the frequency domain. Turning to the emission spectra, our numerical
calculations reveal that when the bias voltage is increased, the emission
antiresonance observed in the regime $|eV|\ll\Gamma$ (see Figs. \ref{figure_2}
and \ref{figure_3}) transforms into a pair antiresonance-resonance located at
the frequency $\hbar\omega=-|eV|/2$. It turns out that in the regime of large
bias voltages, $|eV|\gg\Gamma$, this emission antiresonance-resonance pair is
fully identical to the absorption antiresonance-resonance pair at any value of
the Majorana tunneling phase difference $\Delta\phi$ as demonstrated in
Figs. \ref{figure_6}(a) and (b). Note that both the emission and absorption
antiresonance-resonance pairs are of pure Majorana interference nature because
both of them appear only at finite values of $|\eta_2|$ and $\Delta\phi$
whereas for $|\eta_2|=0$ or $\Delta\phi=0$, when the QD is effectively coupled
to a single MBS, the differential thermoelectric quantum noise is suppressed
at all frequencies for $|eV|\gg\Gamma$. Since the absorption and emission
antiresonance-resonance pairs are identical, we conclude that at large bias
voltages, $|eV|\gg\Gamma$, the Majorana interference makes the differential
thermoelectric quantum noise antisymmetric with respect to the frequency
$\omega$, that is 
\begin{equation}
  \begin{split}
    &\frac{\partial S^>(-\omega,V,V_T,\Delta\phi)}{\partial V_T}=
    -\frac{\partial S^>(\omega,V,V_T,\Delta\phi)}{\partial V_T},\\
    &k_\text{B}T\ll\Gamma,\quad|eV|\gg\Gamma,\quad 0\leqslant\Delta\phi<2\pi.
  \end{split}
\label{Antisymm_DTQN_small_T}
\end{equation}
Thus it is the interplay of the Majorana interference and strong
nonequilibrium which establishes the antisymmetric relation, expressed by
Eq. (\ref{Antisymm_DTQN_small_T}), between the absorption and emission spectra
of the differential thermoelectric quantum noise.

At this point a qualitative physical explanation of the numerical results
would be appropriate. Particularly, it would be useful to try to understand
why the emission and absorption spectra are different at small bias voltages,
$|eV|\ll\Gamma$, and become similar at large bias voltages,
$|eV|\gg\Gamma$. To this end, we note that the behavior of the differential
thermoelectric quantum noise in vicinities of $\hbar\omega=\mp|eV|/2$ is
governed by two tunneling processes. The first one is from the TS to the left
metallic contact and it is this process which is responsible for the emission
spectra arising around $\hbar\omega=-|eV|/2$. The second process is the
tunneling from the left contact to the TS, and it is responsible for the
absorption spectra arising around $\hbar\omega=|eV|/2$. An emission of a
photon with an energy $\hbar\omega\sim|eV|/2$ decreases the energy of a
quasiparticle and it goes from the zero energy in the TS to a negative energy
in a vicinity of the chemical potential of the left contact. An absorption of
a photon with an energy $\hbar\omega\sim|eV|/2$ increases the energy of a
quasiparticle and it goes from a negative energy in a vicinity of the chemical
potential of the left contact to the zero energy in the TS. Note, that we
consider emission or absorption tunneling processes which, respectively, end
in or originate from the vicinity of width $k_\text{B}T$ around the chemical
potential of the left contact, $\mu_L=-|eV|/2$, because we consider the
differential thermoelectric quantum noise. It is given as the derivative over
$V_T$ which is finite only within the vicinity of width $k_\text{B}T$ around
$\hbar\omega=-|eV|/2$ or $\hbar\omega=|eV|/2$ for, respectively, emission or
absorption processes. The difference between the emission and absorption
spectra for the case of small bias voltages, $|eV|\ll\Gamma$, may be
qualitatively understood if we take into account that the two tunneling
processes mentioned above occur via the QD and thus involve the nonequilibrium
distribution function, $n_d(\epsilon)$, of the QD. Since the QD is
noninteracting, its distribution function is of the double step form (see
Ref. \cite{Altland_2010}),
$n_d(\epsilon)=[n_L(\epsilon)+n_R(\epsilon)]/2$. More specifically, since the
differential thermoelectric quantum noise is given by the derivative over
$V_T$, the QD distribution function $n_d(\epsilon)$ contributes to an emission
or absorption process through the thermally broadened region of
$n_L(\epsilon)/2$, that is again in the vicinity of width $k_\text{B}T$ around
$\hbar\omega=-|eV|/2$ or $\hbar\omega=|eV|/2$, respectively. Note, that the
zero energy level of the QD is induced by MBSs and is broadened by the
coupling to the contacts. Its width is given by $\Gamma$. Thus the thermally
broadened region of $n_d(\epsilon)$, namely the vicinity of $\mu_L$ of width
$k_\text{B}T$, is fully within the width $\Gamma$ of the QD zero energy level
when $|eV|\ll\Gamma$ and it participates in an emission and absorption
tunneling process through the factors $n_d(\epsilon)$, $[1-n_d(\epsilon)]$
combined with the factors $n_L(\epsilon)$, $[1-n_L(\epsilon)]$. These
combinations are different for emission and absorption processes which
qualitatively explains the difference between, respectively, the emission and
absorption spectra obtained after integrating these combinations over the
energy domain. Now, turning to the case of large bias voltages,
$|eV|\gg\Gamma$, we note that the derivative of the QD distribution function
$n_d(\epsilon)$ over $V_T$ does not contribute anymore to the differential
thermoelectric quantum noise
$\partial S^>(\omega,V,V_T,\Delta\phi)/\partial V_T$ because the thermally
broadened region of $n_d(\epsilon)$ around $\mu_L$ is fully outside the width
$\Gamma$ of the QD zero energy level when $|eV|\gg\Gamma$. Instead, the
factors $n_d(\epsilon)$, $[1-n_d(\epsilon)]$ become equal,
$n_d(\epsilon)=1-n_d(\epsilon)=1/2$, over the whole width $\Gamma$ of the QD
zero energy level for $|eV|\gg\Gamma$ and their combinations with the factors
$n_L(\epsilon)$, $[1-n_L(\epsilon)]$ result in similar emission and absorption
spectra obtained after integrations in the energy domain. Moreover, since the
contribution from the derivative of $n_d(\epsilon)$ over $V_T$ is similar to
the contribution of the derivative of $n_L(\epsilon)$ over $V_T$, it looks
plausible that the disappearance of the former contribution to the derivative
$\partial S^>(\omega,V,V_T,\Delta\phi)/\partial V_T$ in the regime
$|eV|\gg\Gamma$ reduces the amplitudes of the antiresonance-resonance pairs
twice in comparison with the regime $|eV|\ll\Gamma$.

\begin{figure}
  \includegraphics[width=8.0 cm]{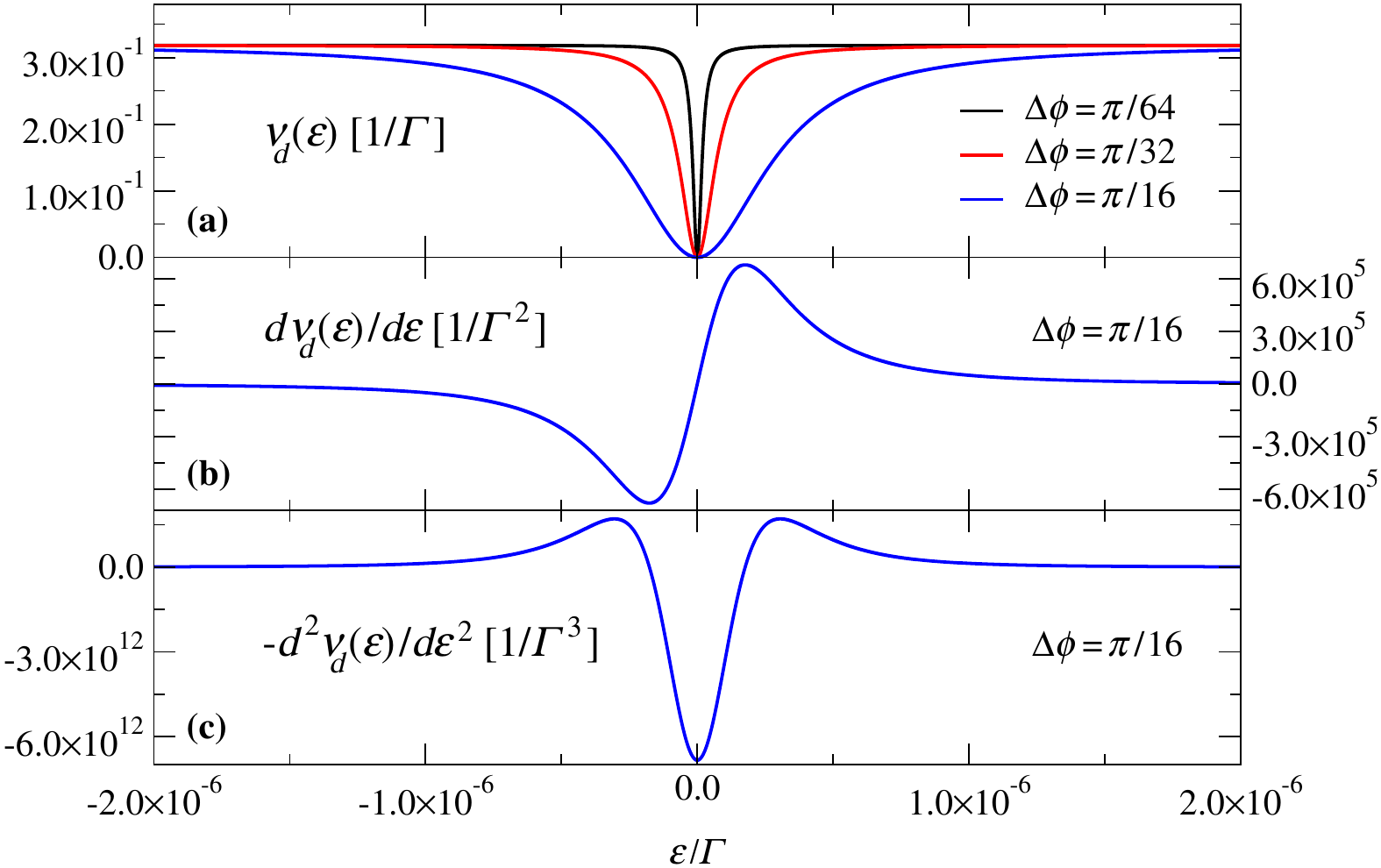}
  \caption{\label{figure_7} Panel ({\bf a}): Density of states of the QD,
    $\nu_d(\epsilon)$, for $\Delta\phi=\pi/64$ (black), $\Delta\phi=\pi/32$
    (red), $\Delta\phi=\pi/16$ (blue). Panel ({\bf b}): First derivative of
    the density of states, $d\nu_d(\epsilon)/d\epsilon$, for
    $\Delta\phi=\pi/16$. Panel ({\bf c}): Second derivative of the density of
    states,$-d^2\nu_d(\epsilon)/d\epsilon^2$, for $\Delta\phi=\pi/16$. In all
    the panels $\epsilon_d/\Gamma=10$, $|\eta_1|/\Gamma=10^3$,
    $|\eta_2|/\Gamma=10^{-3}$, $\xi/\Gamma=10^{-4}$.}
\end{figure}

We would like to emphasize that the explanation given above is rather
qualitative. In particular, it can help to understand why in the regime of
small bias voltages, $|eV|\ll\Gamma$, the emission and absorption spectra are
different and why they are similar in the regime of large bias voltages,
$|eV|\gg\Gamma$, but it cannot explain the exact shape of the spectra. For
example, it does not explain why in the regime of small bias voltages there
develops an antiresonance in the emission spectra and an
antiresonance-resonance pair in the absorption spectra. To understand where
these shapes come from it is important to note that the role of the MBSs is
not only in the appearance of the zero energy level in the QD but also in the
emergence of a strong energy dependence of the density of states
$\nu_d(\epsilon)$ in the QD at low energies. This energy dependence is the
result of the Majorana interference which, on one side, strongly suppresses
the density of states at low energies but, on the other side, it results in
the appearance of wide regions where various derivatives of the density of
states essentially increase. The results presented above reveal that for
$|\eta_2|\neq 0$ and $\Delta\phi\neq 0$ the magnitude of these derivatives and
the size of the regions where they are sufficiently large essentially
determine the behavior of the differential thermoelectric quantum noise
specified as the derivative over $V_T$. In fact, a complex interplay between
the thermoelectric nonequilibrium and Majorana interference arises within and
outside the thermally broadened region of $n_L(\epsilon)$. For example,
outside this region the fluctuation response of quasiparticles to the thermal
voltage $V_T$ would be strongly suppressed for $|\eta_2|=0$ or $\Delta\phi=0$
as expressed by the exponential decay of the derivative of $n_L(\epsilon)$
over $V_T$ outside the domain of width $k_\text{B}T$ around $\mu_L$. However,
for $|\eta_2|\neq 0$ and $\Delta\phi\neq 0$ the region where derivatives of
$\nu_d(\epsilon)$ are sufficiently large may become wider than the thermally
broadened region of $n_L(\epsilon)$. In this situation the corresponding
factors, derivatives of $\nu_d(\epsilon)$ over $\epsilon$ and the derivative
of $n_L(\epsilon)$ over $V_T$, start to compete. It is this competition
between the Majorana interference and thermoelectric nonequilibrium which
results in specific shapes of the emission and absorption spectra within and
outside the thermally broadened region of $n_L(\epsilon)$. To support our
reasoning, we show in Fig. \ref{figure_7} the QD density of states
(Fig. \ref{figure_7}(a)), its first (Fig. \ref{figure_7}(b)) and second
(Fig. \ref{figure_7}(c)) derivatives over the energy. Comparing the shape of
the curves in Fig. \ref{figure_2} with the shape of the curve in
Fig. \ref{figure_7}(c) as well as comparing Fig. \ref{figure_5} with
Fig. \ref{figure_7}(b), we see that the difference between the emission and
absorption spectra is quantitatively rooted in the fact that in the regime of
small bias voltages, $|eV|\ll\Gamma$, the emission spectra are governed by the
second derivative of the QD density of states,
$-d^2\nu_d(\epsilon)/d\epsilon^2$, whereas the absorption spectra are governed
by its first derivative, $d\nu_d(\epsilon)/d\epsilon$. We also see that for
large bias voltages, $|eV|\gg\Gamma$, the emission and absorption spectra
shown in Fig. \ref{figure_6} are both governed by the first derivative
$d\nu_d(\epsilon)/d\epsilon$ of the QD density of states shown in
Fig. \ref{figure_7}(b). Note also, that the derivatives shown in
Figs. \ref{figure_7}(b) and (c) are huge and thus are able to compete with the
exponential decay of the derivative of $n_L(\epsilon)$ over $V_T$ outside the
thermally broadened region to produce values of
$\partial S^>(\omega,V,V_T,\Delta\phi)/\partial V_T$ which would be
sufficiently large for an experimental detection.

Let us recall that the results presented above have been obtained in the
regime specified by $eV_T\ll k_\text{B}T\ll|eV|<|\eta_1|$. Within this regime
we have analyzed the two cases, $|eV|\ll\Gamma$ and $|eV|\gg\Gamma$. So, the
upper bound for the temperature is always the bias voltage. Specifically,
$k_\text{B}T$ should be at least one order of magnitude less than $|eV|$. In
turn, the upper bound for the bias voltage is always $|\eta_1|$. The above two
cases are characterized by low temperatures, $k_\text{B}T\ll\Gamma$. The third
case within the regime $eV_T\ll k_\text{B}T\ll|eV|<|\eta_1|$ is specified by
high temperatures, $k_\text{B}T\gg\Gamma$, and is analyzed below.
\begin{figure}
  \includegraphics[width=8.0 cm]{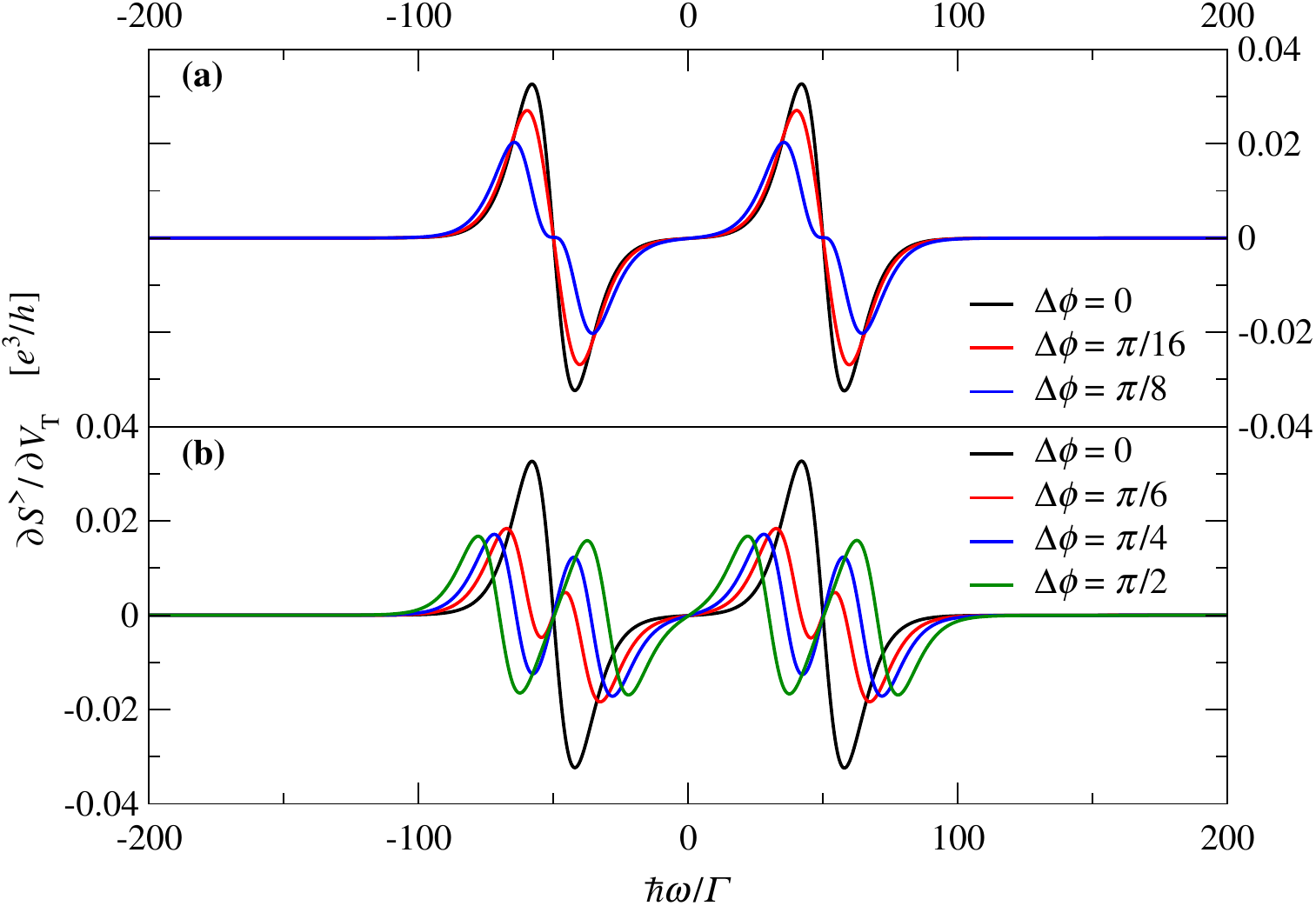}
  \caption{\label{figure_8} Frequency dependence of the differential
    thermoelectric quantum noise,
    $\partial S^>(\omega,V,V_T,\Delta\phi)/\partial V_T$, in the emission
    ($\omega<0$) and absorption ($\omega>0$) domains for large bias voltages and
    temperatures, $|eV|\gg\Gamma$, $k_\text{B}T\gg\Gamma$, and for various
    values of the Majorana tunneling phase difference
    $\Delta\phi$. Specifically, for all the curves $|eV|/\Gamma=10^2$,
    $k_\text{B}T/\Gamma=5.0$ and, in Panel ({\bf a}), $\Delta\phi=0$ (black),
    $\Delta\phi=\pi/16$ (red), $\Delta\phi=\pi/8$ (blue), in Panel ({\bf b}),
    $\Delta\phi=0$ (black), $\Delta\phi=\pi/6$ (red), $\Delta\phi=\pi/4$ (blue),
    $\Delta\phi=\pi/2$ (green). The values of the other parameters:
    $\epsilon_d/\Gamma=10$, $eV_T/\Gamma=10^{-3}$, $|\eta_1|/\Gamma=2.5\times
    10^2$, $|\eta_2|/\Gamma=10$, $\xi/\Gamma=10^{-4}$.}
\end{figure}

It turns out that the antisymmetric relation between the absorption and
emission spectra, Eq. (\ref{Antisymm_DTQN_small_T}), arises in strong
nonequilibrium also when the QD is effectively coupled to a single MBS, that
is when $|\eta_2|=0$ or $\Delta\phi=0$. This happens in the regime of high
temperatures, $k_\text{B}T\gg\Gamma$, as demonstrated by the black curve in
Fig. \ref{figure_8}(a) or (b) for which $\Delta\phi=0$. In this case the two
MBSs do not interfere with different tunneling phases but
$\partial S^>(\omega,V,V_T,\Delta\phi)/\partial V_T$ is not suppressed at all
frequencies as has been observed above in the regime of low temperatures,
$k_\text{B}T\ll\Gamma$. Indeed, at high temperatures there arise thermally
excited Majorana resonance-antiresonance pairs. They have a non-interfering
nature and are located at $\hbar\omega=\pm|eV|/2$ in the absorption and
emission spectra, respectively. When the interference of the two MBSs with
different tunneling phases is gradually switched on by increasing $\Delta\phi$
at a finite value of $|\eta_2|$, the amplitudes of the thermal resonance and
antiresonance start to decrease as can be seen from the red curve with
$\Delta\phi=\pi/16$ in Fig. \ref{figure_8}(a). Further increase of
$\Delta\phi$ produces a notable reduction of the negative slopes in small
vicinities of $\hbar\omega=\pm|eV|/2$ (blue curve with $\Delta\phi=\pi/8$ in
Fig. \ref{figure_8}(a)) which is a precursor to the formation of the Majorana
interference induced antiresonance-resonance pairs. For larger values of
$\Delta\phi$ the Majorana interference pattern fully develops. Indeed, as
demonstrated in Fig. \ref{figure_8}(b), in the centers of the thermally
excited pairs resonance-antiresonance there emerge pairs
\begin{figure}
\includegraphics[width=8.0 cm]{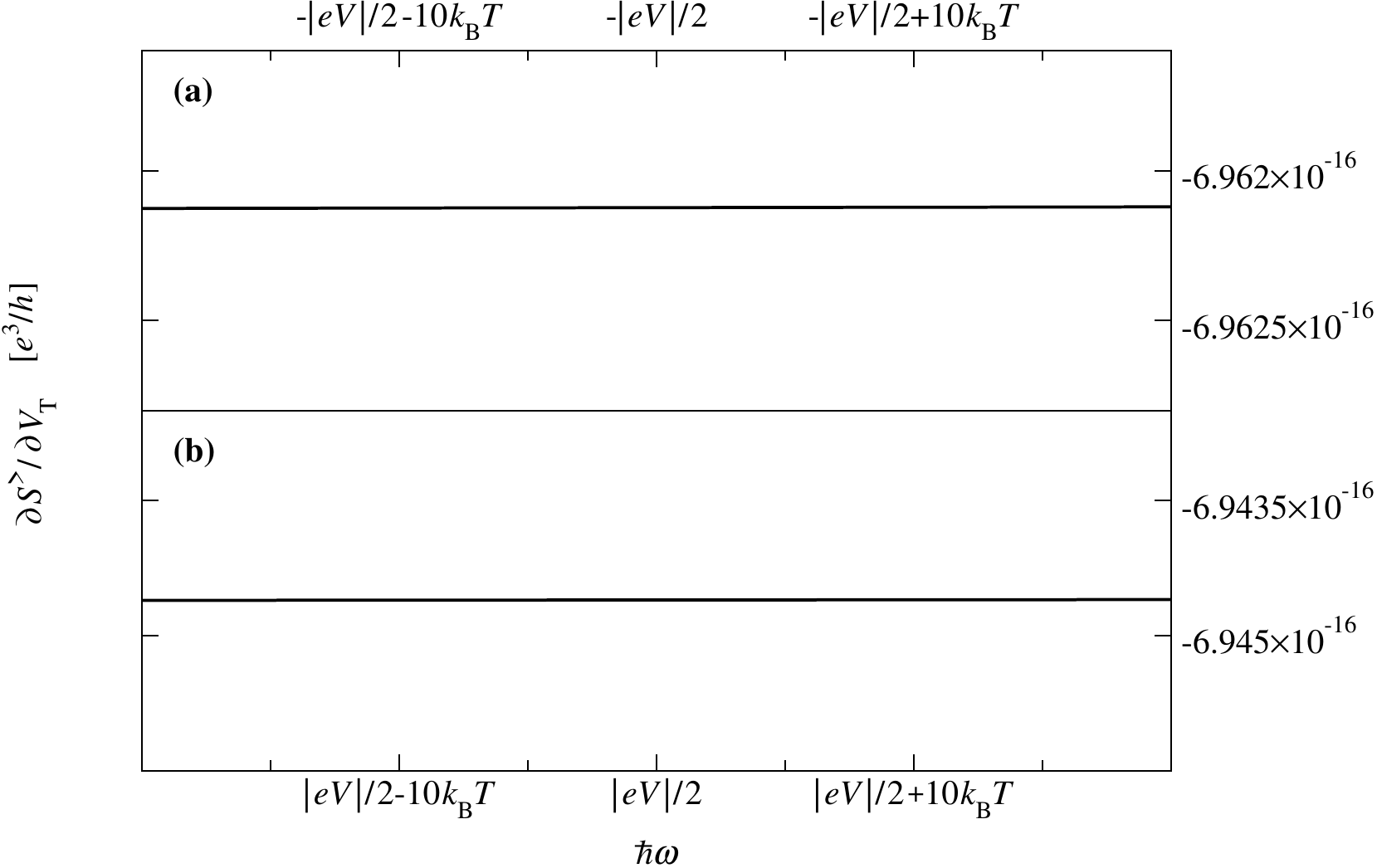}
\caption{\label{figure_9} Frequency dependence of the differential
  thermoelectric quantum noise,
  $\partial S^>(\omega,V,V_T,\Delta\phi)/\partial V_T$, for ABSs. Panels
  ({\bf a}) and ({\bf b}) show, respectively, the emission ($\omega<0$) and
  absorption ($\omega>0$) domains in the vicinities of $\hbar\omega=\mp|eV|/2$
  in the same regime of small temperatures and bias voltages as the one used
  for Figs. \ref{figure_2} and \ref{figure_5} for the case of MBSs. Here
  $|\eta_2|/\Gamma=5.0\times 10^2$, $\xi/\Gamma=10^2$ and
  $\Delta\phi=\pi/16$. The values of the other parameters are the same as in
  Fig. \ref{figure_2}.}
\end{figure}
antiresonance-resonance induced by the Majorana interference. This is clearly
seen by comparing the black curve, corresponding to the system with the QD
effectively coupled to a single MBS ($\Delta\phi=0$), with the red
($\Delta\phi=\pi/6$), blue ($\Delta\phi=\pi/4$) and green ($\Delta\phi=\pi/2$)
curves, corresponding to the system where the two MBSs interfere with
different tunneling phases. Note, that for $\Delta\phi=\pi/2$ the Majorana
interference becomes so strong that the amplitudes of the antiresonance and
resonance in the Majorana interference induced pairs antiresonance-resonance
are almost equal to the ones in the thermally excited pairs
resonance-antiresonance. What is important to note is that the Majorana
interference does not break the original (at $\Delta\phi=0$) antisymmetric
character of the differential thermoelectric quantum noise composed in the
absence of the Majorana interference of the two thermally excited pairs
resonance-antiresonance. Indeed, the absorption and emission
antiresonance-resonance pairs induced by the Majorana interference in the
centers of the corresponding thermally excited resonance-antiresonance pairs
are fully identical. Therefore, in the regime of high temperatures,
$k_\text{B}T\gg\Gamma$, the differential thermoelectric quantum noise is also
antisymmetric,
\begin{equation}
  \begin{split}
    &\frac{\partial S^>(-\omega,V,V_T,\Delta\phi)}{\partial V_T}=
    -\frac{\partial S^>(\omega,V,V_T,\Delta\phi)}{\partial V_T},\\
    &k_\text{B}T\gg\Gamma,\quad|eV|\gg\Gamma,\quad 0\leqslant\Delta\phi<2\pi.
  \end{split}
\label{Antisymm_DTQN_large_T}
\end{equation}
However, in this case the antisymmetry is produced by an interplay of highly
nonequilibrium states and thermal fluctuations whereas the Majorana
interference, when sufficiently strong, introduces characteristic
antiresonance-resonance pairs without breaking the antisymmetric relation
between the absorption and emission spectra. The picture described above
qualitatively changes when MBSs are replaced with ABSs as discussed below.
\begin{figure}
\includegraphics[width=8.0 cm]{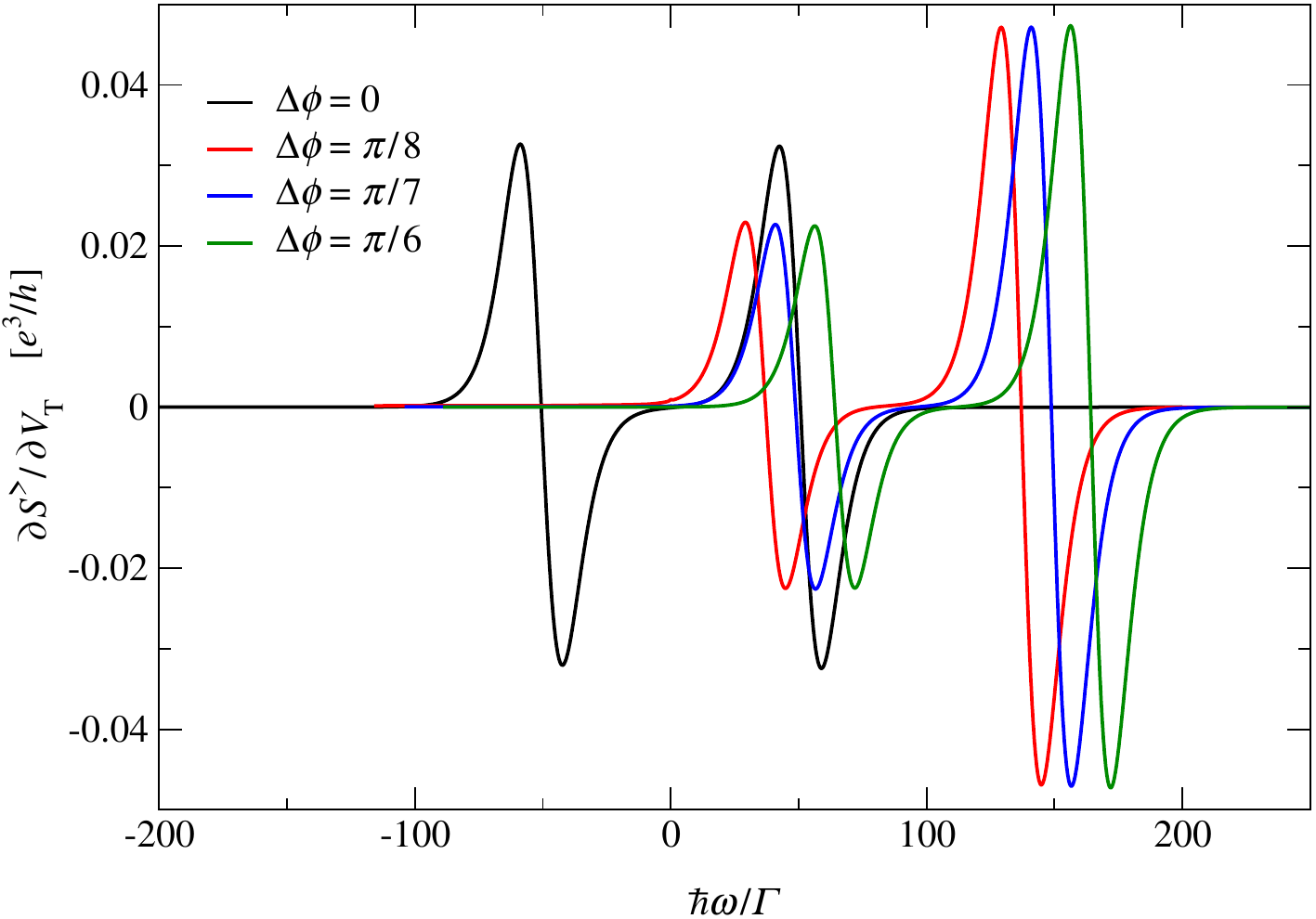}
\caption{\label{figure_10} Frequency dependence of the differential
  thermoelectric quantum noise,
  $\partial S^>(\omega,V,V_T,\Delta\phi)/\partial V_T$, for ABSs. The emission
  ($\omega<0$) and absorption ($\omega>0$) domains are shown in the regime of
  large bias voltages and temperatures, $|eV|\gg\Gamma$,
  $k_\text{B}T\gg\Gamma$, for various values of the tunneling phase difference
  $\Delta\phi$. Specifically, for all the curves $|eV|/\Gamma=10^2$,
  $k_\text{B}T/\Gamma=5.0$ and $\Delta\phi=0$ (black), $\Delta\phi=\pi/8$
  (red), $\Delta\phi=\pi/7$ (blue), $\Delta\phi=\pi/6$ (green). The values of
  the other parameters: $\epsilon_d/\Gamma=10$, $eV_T/\Gamma=10^{-3}$,
  $|\eta_1|/\Gamma=2.5\times 10^2$, $|\eta_2|/\Gamma=1.25\times 10^2$,
  $\xi/\Gamma=10^2$.}
\end{figure}

To analyze the differential thermoelectric quantum noise resulting from ABSs
coupled to the QD and compare its behavior with the one discussed above for
the QD interacting with MBSs one may apply our model in a suitable
regime. Namely, the QD interacting with ABSs may be modeled (see
Refs. \cite{Deng_2018,Smirnov_2024}) by using a large value of the overlap
energy $\xi$ and a value of the second Majorana tunneling amplitude $|\eta_2|$
of the same order as the value of the first one, that is
$|\eta_2|\sim|\eta_1|$. In the low temperature regime both for small
($|eV|\ll\Gamma$) and large ($|eV|\gg\Gamma$) bias voltages the differential
thermoelectric quantum noise induced by ABSs turns out to be strongly
suppressed in vicinities of the frequencies $\hbar\omega=\mp|eV|/2$. For
example, Figs. \ref{figure_9}(a) and (b) show
$\partial S^>(\omega,V,V_T,\Delta\phi)/\partial V_T$ for ABSs in the
vicinities of, respectively, $\hbar\omega=\mp|eV|/2$ for $|eV|\ll\Gamma$,
where in the case of MBSs there appear the emission antiresonance shown in
Fig. \ref{figure_2} in the vicinity of the frequency $\hbar\omega=-|eV|/2$ and
the absorption antiresonance-resonance pair shown in Fig. \ref{figure_5} in
the vicinity of the frequency $\hbar\omega=|eV|/2$. As we can see from
Figs. \ref{figure_9}(a) and (b), the differential thermoelectric quantum noise
induced by ABSs in the regime of low temperatures and small bias voltages is
suppressed several orders of magnitude in comparison with the case of MBSs and
does not exhibit any emission antiresonance and absorption
antiresonance-resonance pair. Additionally, within the low temperature regime,
the differential thermoelectric quantum noise induced by ABSs turns out to be
strongly suppressed around the frequencies $\hbar\omega=\mp|eV|/2$ also for
large bias voltages, $|eV|\gg\Gamma$, and, in contrast to the case of MBSs
(see Fig. \ref{figure_6}), does not exhibit any antiresonance-resonance pair
for any value of $\Delta\phi$. In the high temperature regime the behavior of
the differential thermoelectric quantum noise induced by ABSs is shown in
Fig. \ref{figure_10}. Comparing it with the one discussed above for MBSs in
Fig. \ref{figure_8} one explicitly observes an exceptional role of the
Majorana interference. Indeed, as we can see, for $\Delta\phi=0$ the black
curve in Fig. \ref{figure_8}(a) or (b) and the black curve in
Fig. \ref{figure_10} are identical. This means that when the QD is effectively
coupled to a single MBS, the differential thermoelectric quantum noise in the
regime of large bias voltages and temperatures, $|eV|\gg\Gamma$,
$k_\text{B}T\gg\Gamma$, would not be able to distinguish this single MBS from
ABSs. Remarkably, when $\Delta\phi\neq 0$, interference effects reveal a
fundamental difference between fluctuation responses of the two MBSs with
different tunneling phases and ABSs making the differential thermoelectric
quantum noise of ABSs qualitatively different from the one characterizing
MBSs. Indeed, as one can see in Fig. \ref{figure_10}, finite values of
$\Delta\phi$ do not produce any pairs antiresonance-resonance in the centers
of the thermally excited resonance-antiresonance pairs as it happens for the
system with the MBSs (see Fig. \ref{figure_8}). This is clearly seen in the
red, blue and green curves with $\Delta\phi=\pi/8$, $\Delta\phi=\pi/7$ and
$\Delta\phi=\pi/6$, respectively. Instead, Fig. \ref{figure_10} shows that the
thermally excited pairs resonance-antiresonance shift to positive frequencies,
that is to the absorption spectra. Additionally, the amplitudes of the
resonance and antiresonance in the pair located at a larger frequency increase
whereas those in the pair located at a smaller frequency decrease. As a
result, we see that for ABSs the differential thermoelectric quantum noise
loses its antisymmetry as a function of the frequency $\omega$, that is
\begin{equation}
  \begin{split}
    \frac{\partial S^>(-\omega,V,V_T,\Delta\phi)}{\partial V_T}&\neq
    -\frac{\partial S^>(\omega,V,V_T,\Delta\phi)}{\partial V_T},\\
    \Delta\phi&\neq 0.
  \end{split}
\label{No_Antisymm_DTQN_ABS}
\end{equation}
Therefore, we conclude that in the high temperature regime exactly the
interference behavior of the differential thermoelectric quantum noise,
emerging for $\Delta\phi\neq 0$, enables one to explicitly and qualitatively
distinguish systems with MBSs from those where fluctuations of thermoelectric
currents are governed by ABSs.
\section{Conclusion}\label{concl}
In conclusion, we have analyzed fluctuations of thermoelectric currents
flowing through a nonequilibrium QD interacting with MBSs. Particularly, we
have focused on the differential thermoelectric quantum noise,
$\partial S^>(\omega,V,V_T,\Delta\phi)/\partial V_T$, which results from an
interplay between  Majorana interference effects induced by the phase
difference $\Delta\phi$ and nonequilibrium states produced by both electric
and thermal voltages, $V$ and $V_T$, respectively, at finite frequencies
$\omega$. In the regime of low bias voltages we have found an emission
($\omega<0$) antiresonance located at $\hbar\omega=-|eV|/2$. On the absorption
($\omega>0$) side of the spectra we have found an antiresonance-resonance pair
with its center at $\hbar\omega=|eV|/2$. The minimum and maximum of the
antiresonance and resonance are symmetrically located with respect to the
frequency $\hbar\omega=|eV|/2$ and have equal amplitudes. Both the emission
antiresonance and absorption antiresonance-resonance pair involve the Majorana
modes with $\Delta\phi\neq 0$ and thus turn out to be of pure Majorana
interference nature. In the regime of large bias voltages it has been found
that the absorption antiresonance-resonance pair persists but the amplitudes
of the antiresonance and resonance are reduced exactly twice. At the same
time, the emission antiresonance disappears in the regime of large bias
voltages and turns into an antiresonance-resonance pair. This emission
antiresonance-resonance pair is centered at $\hbar\omega=-|eV|/2$ and is fully
identical to the absorption one. As a result, the differential thermoelectric
quantum noise becomes antisymmetric as a function of the frequency $\omega$,
that is the absorption and emission spectra are related as
$\partial S^>(-\omega,V,V_T,\Delta\phi)/\partial V_T=
-\partial S^>(\omega,V,V_T,\Delta\phi)/\partial V_T$. This is in contrast to
the regime of low bias voltages where the absorption and emission spectra are
in general not related to each other. Finally, we have analyzed the regime of
large bias voltages and temperatures. It has been demonstrated that in this
regime the antisymmetric relation between the absorption and emission spectra
arises even when the QD is effectively coupled to a single MBS, that is at
$\Delta\phi=0$. However, when $\Delta\phi\neq 0$, the differential
thermoelectric quantum noise acquires characteristic interference patterns in
the absorption and emission spectra in such a way that the antisymmetric
relation between these spectra does not break. This is in contrast to the
system where the QD interacts with ABSs. In this system the differential
thermoelectric quantum noise induced by ABSs is strongly suppressed in the low
temperature regime both for small and large bias voltages whereas in the high
temperature regime it does not exhibit any interference pattern for
$\Delta\phi\neq 0$ and is only characterized by a simple shift to positive
frequencies, that is to the absorption side, which obviously breaks the
antisymmetric relation between the absorption and emission spectra.

The regime of large bias voltages and temperatures might be attractive for
experiments. Let us estimate the bias voltages $|eV|$ and temperatures
$k_\text{B}T$ at which experiments could be performed and compared with the
results shown in Figs. \ref{figure_8} and \ref{figure_10} for the systems with
MBSs and ABSs, respectively. As specified above, the largest energy scale in
the problem is $|\eta_1|$ which is thus the upper bound for the bias
voltage, $|eV|<|\eta_1|$. Taking into account that the largest energy scale
must not exceed the induced superconducting gap $\Delta$, we have
$|\eta_1|\lesssim\Delta$. In Figs. \ref{figure_8} and \ref{figure_10} we have
$k_\text{B}T=2.0\times 10^{-2}|\eta_1|$ which means that
$k_\text{B}T\lesssim 2.0\times 10^{-2}\Delta$. Taking the value of $\Delta$
from a typical experiment, for example from Ref. \cite{Mourik_2012},
$\Delta\approx 250\,\mu\text{eV}$, one obtains estimates for the bias voltage
and thermal energy: $|eV|<250\,\mu\text{eV}$ and
$k_\text{B}T\lesssim 5\,\mu\text{eV}$. In terms of the temperature the latter
estimate is equivalent to $T\lesssim 60\,\text{mK}$. Therefore, the upper
bounds for the bias voltages and temperatures corresponding to the results
presented in Figs. \ref{figure_8} and \ref{figure_10} are about
$250\,\mu\text{eV}$ and $60\,\text{mK}$, respectively. Since such bias
voltages and temperatures are quite feasible, we hope that our results might
be attractive for experimental measurements of the differential thermoelectric
quantum noise of interfering MBSs at finite frequencies.
\section*{Acknowledgments}
The author thanks Reinhold Egger for a very useful discussion.

\end{document}